\documentclass[aps,prl,reprint,superscriptaddress]{revtex4-1}

\usepackage{blindtext}
\usepackage{mhchem}
\usepackage{hyperref}
\usepackage{bm}
\usepackage{siunitx}
\usepackage{mathrsfs}

\begin{document}


\title{Second Harmonic Generation as a Probe of Broken Mirror Symmetry}


\author{Bryan T. Fichera}
\affiliation{Department of Physics, Massachusetts Institute of Technology, Cambridge, Massachusetts 02139, USA}
\author{Anshul Kogar}
\affiliation{Department of Physics, Massachusetts Institute of Technology, Cambridge, Massachusetts 02139, USA}
\author{Linda Ye}
\affiliation{Department of Physics, Massachusetts Institute of Technology, Cambridge, Massachusetts 02139, USA}
\author{Bilal G{\"o}kce}
\affiliation{Department of Physics, Massachusetts Institute of Technology, Cambridge, Massachusetts 02139, USA}
\affiliation{Technical Chemistry I and Center for Nanointegration Duisburg-Essen (CENIDE), University of Duisburg-Essen, 45141 Essen, Germany}
\author{Alfred Zong}
\affiliation{Department of Physics, Massachusetts Institute of Technology, Cambridge, Massachusetts 02139, USA}
\author{Joseph G. Checkelsky}
\affiliation{Department of Physics, Massachusetts Institute of Technology, Cambridge, Massachusetts 02139, USA}
\author{Nuh Gedik}
\email[]{gedik@mit.edu}
\affiliation{Department of Physics, Massachusetts Institute of Technology, Cambridge, Massachusetts 02139, USA}


\date{\today}

\begin{abstract}
The notion of spontaneous symmetry breaking has been used to describe phase transitions in a variety of physical systems.
In crystalline solids, the breaking of certain symmetries, such as mirror symmetry, is difficult to detect unambiguously.
Using 1$T$-\ce{TaS2}, we demonstrate here that rotational-anisotropy second harmonic generation (RA-SHG) is not only a sensitive technique for the detection of broken mirror symmetry, but also that it can differentiate between mirror symmetry-broken structures of opposite planar chirality.
We also show that our analysis is applicable to a wide class of different materials with mirror symmetry-breaking transitions.
Lastly, we find evidence for bulk mirror symmetry-breaking in the incommensurate charge density wave phase of 1$T$-\ce{TaS2}.
Our results pave the way for RA-SHG to probe candidate materials where broken mirror symmetry may play a pivotal role.
\end{abstract}


\maketitle

In condensed matter systems, phases are often classified by the symmetries that they break.
Identifying these symmetries enables one to understand a system's order parameter, collective excitations, topological defects, and allowable topological indices \cite{sethna, thouless}.
Together, these attributes allow one to predict how a material will respond to external perturbations like electromagnetic fields, heat, and mechanical forces, which is a central goal of condensed matter physics.

Specifically, the presence or absence of mirror symmetry can lead to a variety of unusual phases and properties.
For example, the absence of mirror plane symmetry in noncentrosymmetric materials can give rise to gyrotropic order, which can lead to a nonzero out-of-plane circular photo-galvanic effect \cite{belinicher1980}.
Moreover, in certain topological crystalline insulators, such as SnTe, the presence of mirror symmetry can give rise to conducting surface states through the existence of a nonzero mirror Chern number.
This topological index guarantees an even number of Dirac cones on surfaces where mirror symmetry is retained \cite{teo2008, hsieh_topological_2012, fu_topological_2011}.

%
In other circumstances, like that in the pseudogap regime of cuprate superconductors, the existence of mirror symmetry is more controversial and has led some to seek an experimental method to serve as a binary indicator of broken mirror symmetry \cite{hlobil_elastoconductivity_2015}.
In principle, several tools can already do this, including resonant ultrasound spectroscopy \cite{leisure1997, migliori1993}, X-ray, neutron, and electron diffraction \cite{bacon1966, dorset2013}, and a recently-proposed method, shear conductivity \cite{hlobil_elastoconductivity_2015}.
Resonant ultrasound spectroscopy and the diffraction-based techniques are more sensitive to the ionic lattice, which makes the identification of subtle electronic symmetry-breaking challenging.
And while shear conductivity has the potential to be an extremely versatile tool for identifying broken point group symmetries, experimental pursuits are currently only preliminary \cite{hlobil_elastoconductivity_2015}.
In this study, we focus on a nonlinear optical technique, rotational-anisotropy second harmonic generation (RA-SHG), which we show is capable not only of identifying broken mirror symmetry \cite{heinz_study_1985} but also of resolving its sense (left- or right-handed).
Furthermore, RA-SHG is sensitive to the electronic subsystem and can be used for microscopy studies, making it an ideal experimental tool for probing phase transitions where domains may arise \cite{kumar_magnetic_2017, harter_parity-breaking_2017}.

In this direction, we choose a material, 1$T$-\ce{TaS2}, in which vertical mirror plane symmetry is manifestly broken across an incommensurate (IC) to nearly commensurate (NC) charge density wave (CDW) transition \cite{wilson_charge-density_1975, zong_ultrafast_2018}.
Using this material, we demonstrate in this Rapid Communication that RA-SHG is an effective probe of broken mirror symmetry.
We also show that the sense (i.e. left- or right-handed) associated with the mirror symmetry-broken structure is encoded in the angular dependence of the RA-SHG signal.
Thus, RA-SHG can be used to differentiate mirror-opposite domains. 
While the data presented in this work is specific to the case of 1$T$-\ce{TaS2}, we also show analytically that our technique is applicable to a wide class of transitions involving spontaneously broken mirror symmetry.

\begin{figure}
\includegraphics{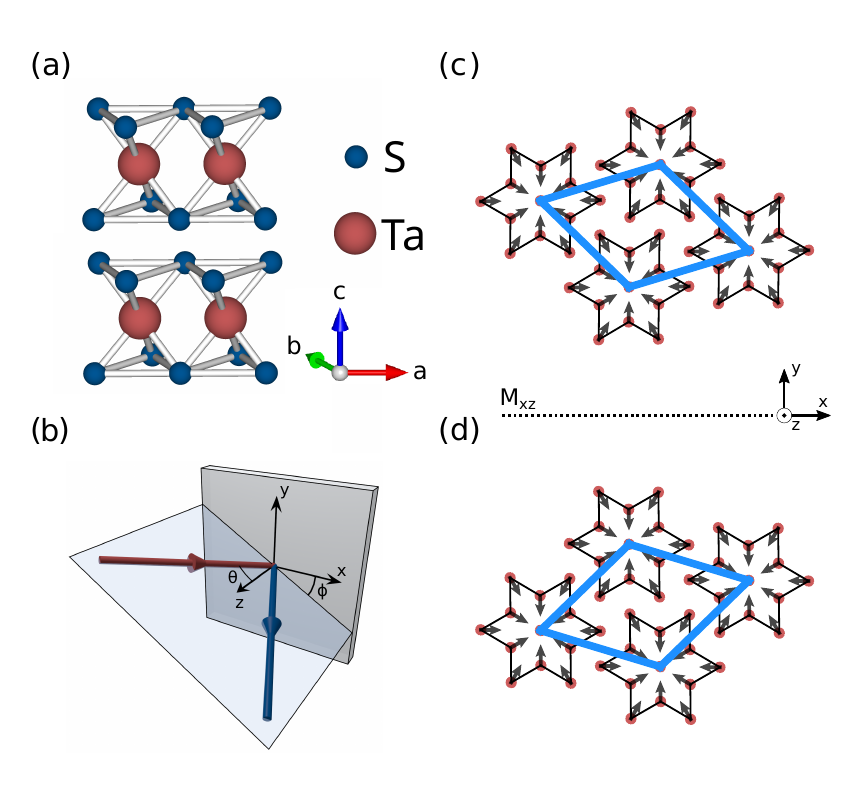}%
\caption{\label{fig:fig0}
(a) Structure of 1$T$-\ce{TaS2} in the undistorted phase. 
Ta and S atoms are depicted in red and blue, respectively. 
(b) Schematic of the experimental geometry.
(c-d) Structure of the CDW in the NC phase.
Arrows denote the movement of the Ta atoms (red) below $T_\mathrm{IC-NC}=353\si{K}$ from their undistorted positions.
The transition at $T_{\mathrm{IC-NC}}$ spontaneously breaks mirror symmetry, so that two different CDW configurations ($\alpha$ and $\beta$) can form which have opposite planar chirality.
The new unit cells of the two configurations are depicted in blue.
$M_{xz}$ denotes one vertical mirror plane which is broken beneath $T_\mathrm{IC-NC}$.
The others are related to $M_{xz}$ by $\pm 120^\circ$ rotation about the $z$ axis.}
\end{figure}

1$T$-\ce{TaS2} is a layered material with a crystallographic structure identical to other octahedrally-coordinated transition metal dichalcogenides (Fig.~\ref{fig:fig0}(a)).
The space group of the high temperature, undistorted phase is $P\bar{3}m1$ (no.164, point group $D_{3d}$)~\cite{spijkerman_x-ray_1997}, and the point group of the surface normal to the (001) direction in this phase is $C_{3v}$~\cite{scruby_role_1975, fung_application_1980}.
Upon lowering temperature, 1$T$-\ce{TaS2} undergoes a series of CDW transitions.
At $T_{\mathrm{IC}}=550\si{K}$, a triple-$q$ IC CDW forms which breaks translational symmetry but retains the surface point group symmetries of the undistorted phase~\cite{scruby_role_1975, fung_application_1980}.
The effects of the CDW on the bulk symmetries in this phase are not yet understood.
On further cooling, at $T_{\mathrm{IC-NC}}=353\si{K}$ there is a weak first-order transition to an NC CDW phase where three vertical mirror plane symmetries are broken~\cite{spijkerman_x-ray_1997} and the surface point group becomes $C_3$.
The NC phase has been visualized with scanning tunneling microscopy and exhibits patches of commensurate ``Star of David" hexagrams that are separated by a network of discommensurations \cite{wu_hexagonal_1989}.
Because mirror symmetry is broken, there are two energetically equivalent CDW configurations ($\alpha$ and $\beta$) in the NC phase that have opposite planar chirality (Fig.~\ref{fig:fig0}(c)-(d)).
At even lower temperatures, $T_{\mathrm{NC-C}}=184\si{K}$, 1$T$-\ce{TaS2} undergoes a symmetry-preserving first-order transition, where the discommensurations disappear and the CDW locks into a structure commensurate with the underlying lattice~\cite{ishiguro_high-resolution_1995}.

Recent interest in the NC phase of 1$T$-\ce{TaS2} has arisen due to the possibility of injecting mirror-opposite domains into the CDW structure \cite{zong_ultrafast_2018} which do not otherwise develop during the IC-NC transition.
Zong \textit{et al.} were able to induce these domains using a single ultrafast pulse of light, which was found to drive the material into a long-lived metastable state possessing domains of opposite planar chirality.
Partly motivated by the desire to image these domains, we seek here a simple experimental method that could identify domains with opposite planar chirality.

1$T$-\ce{TaS2} samples used in the experiment were grown using the chemical vapor transport technique, as described in Ref.~\cite{zong_ultrafast_2018}.
We verified that the NC phase of 1$T$-\ce{TaS2} was single-domain by performing electron diffraction on a sample from the same batch\cite{supplementary_materials}.
This is in agreement with previous works~\cite{zong_ultrafast_2018, wilson_charge-density_1975, bovet_pseudogapped_2004, shiba_phenomenological_1986}.

\begin{figure}
\includegraphics{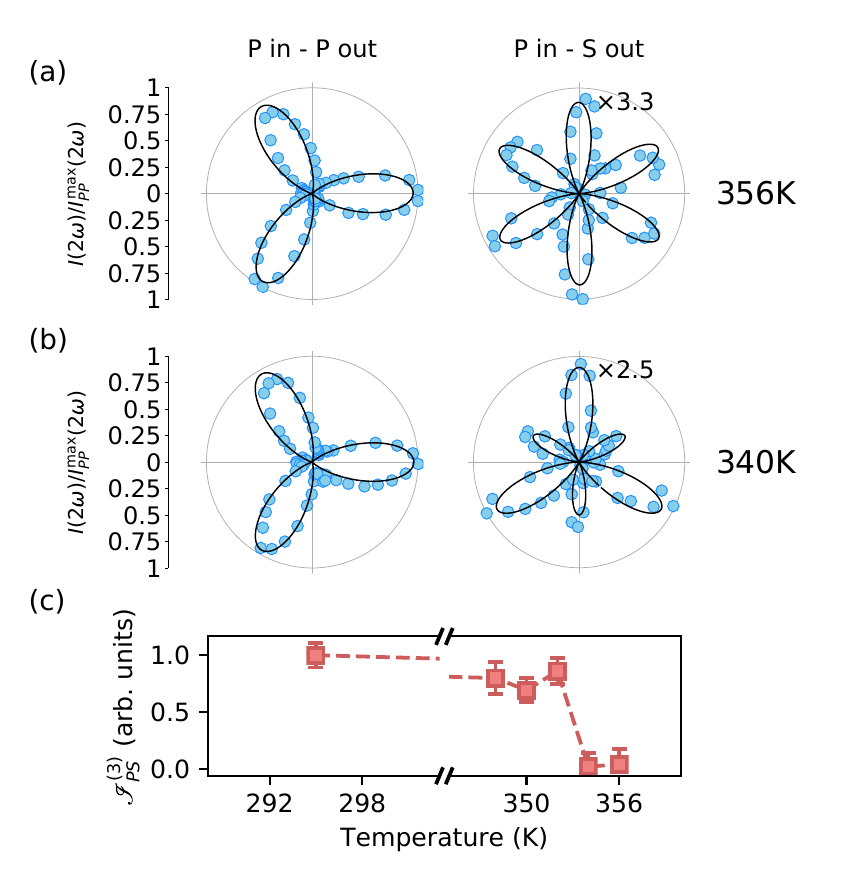}%
\caption{\label{fig:fig1}
(a-b) Second harmonic intensity from 1T-\ce{TaS2} as a function of $\phi$ above (a) and below (b) $T_{\mathrm{IC-NC}}=353\si{K}$.
For clarity, only two polarization channels are shown; the others are reproduced in the supplement\cite{supplementary_materials}.
Solid lines in (a) and (b) are best fits to the data using the surface point groups $C_{3v}$ and $C_3$, respectively.
Data is normalized to the maximum value of the P$_\mathrm{in}$-P$_\mathrm{out}$ signal for each temperature.
(c) Temperature dependence of the third Fourier coefficient, $\mathscr{I}_{PS}^{(3)}$, of the intensity (see main text and supplement\cite{supplementary_materials} for details).
}
\end{figure}

In RA-SHG~\cite{harter_high-speed_2015, lu_fast_2018, torchinsky_structural_2015, lu_fourier_2018}, a pulsed laser beam of frequency $\omega$ and amplitude $E(\omega)$ is focused onto a sample with nonzero angle of incidence $\theta$ (Fig.~\ref{fig:fig0}(b)).
The $2\omega$ component of the radiation emitted by the sample is subsequently measured in various combinations of incoming and outgoing polarizations (either parallel (P) or perpendicular (S) to the plane of incidence) and as a function of the angle $\phi$ between the plane of incidence and some crystallographic axis.
In noncentrosymmetric materials, the response is dominated by the bulk electric dipole moment $P_i(2\omega) = \chi_{ijk}E_j(\omega)E_k(\omega)$ \cite{boyd}, where $\chi_{ijk}$ is a material-specific susceptibility tensor which must be invariant under all symmetry operations present in the crystallographic point group.
In centrosymmetric crystals such as 1$T$-\ce{TaS2}, the bulk electric dipole response is forbidden~\cite{boyd, powell}.
In this case, the dominant response often comes from the surface of the sample, which necessarily breaks inversion symmetry~\cite{bloembergen_optical_1968}.
SHG from surfaces of materials is described by a different susceptibility tensor, $\chi_{ijk}^S$, which is constrained by the crystal symmetries of the surface.
In addition, there can be bulk contributions from higher-order processes which are allowed in the presence of inversion symmetry, such as the bulk quadrupole response $Q_{ij}(2\omega) = \chi^Q_{ijkl}E_k(\omega)E_l(\omega)$~\cite{kumar_magnetic_2017, shen}.
Our experimental implementation uses a fast-rotation setup similar to that described in Refs.~\onlinecite{harter_high-speed_2015} and \onlinecite{torchinsky_low_2014}.
Here, we use an 800\si{nm} (1.55\si{eV}) laser beam incident at $10^\circ$ with respect to the (001) sample normal.
Further experimental details can be found in the supplement\cite{supplementary_materials}.



To show that we are sensitive to the breaking of mirror symmetry across the IC to NC transition, we took RA-SHG data on 1$T$-\ce{TaS2} in both phases.
Figs.~\ref{fig:fig1}(a) and \ref{fig:fig1}(b) show the second harmonic response from the sample above and below $T_{\mathrm{IC-NC}} = 353\si{K}$ in two polarization channels, plotted as a function of $\phi$.
We are able to fit the rotational anisotropy in the IC phase using the surface point group $C_{3v}$ (Fig.~\ref{fig:fig1}(a)), which is in agreement with previous reports~\cite{scruby_role_1975, fung_application_1980}.
It should be noted that in order to fit the S$_\mathrm{in}$-S$_\mathrm{out}$ polarization channel appropriately, we find that it is necessary to add an additional bulk quadrupole contribution to the signal.
The consequences of this contribution will be discussed later, but at the moment they do not affect our conclusions.

\begin{figure}
\includegraphics{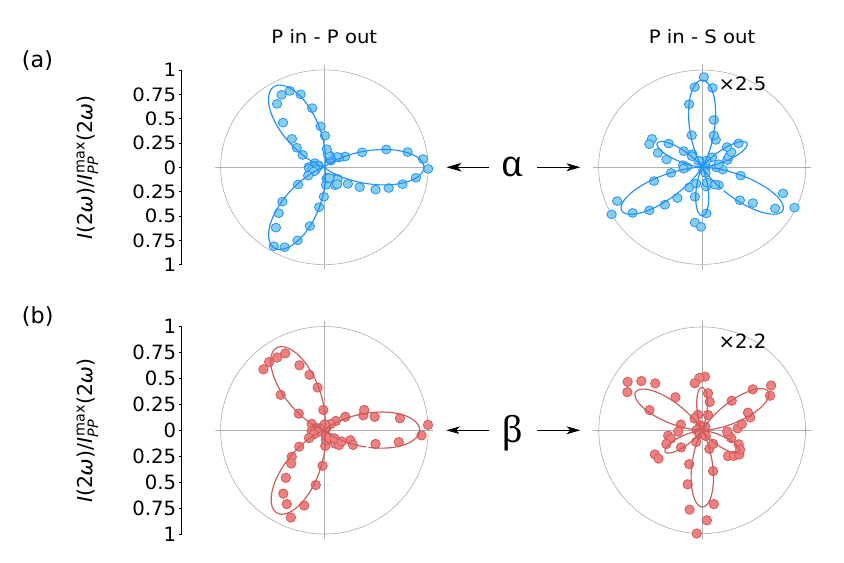}
\caption{\label{fig:fig2}Second harmonic intensity as a function of $\phi$ from mirror-opposite samples of 1$T$-\ce{TaS2} in the NC phase ($T=340\si{K}$).
The labels $\alpha$ and $\beta$ refer to the two degenerate mirror-image configurations which are allowed in the NC phase.
The solid line in (a) is a fit to the data using the surface point group $C_3$.
The fit in (b) was generated by performing a mirror operation\cite{supplementary_materials} to the numerical susceptibility tensor obtained from (a).
Data is normalized to the maximum value in the P$_\mathrm{in}$-P$_\mathrm{out}$ polarization channel for each sample.}
\end{figure}

Upon cooling into the NC phase, mirror symmetry is spontaneously broken and the surface point group reduces to $C_3$~\cite{spijkerman_x-ray_1997}.
As a result, the RA-SHG exhibits a marked lowering of symmetry (Fig.~\ref{fig:fig1}(b)).
This lowering of symmetry can be understood by noting that the $\phi$ dependence of $I_\mathrm{PS}(2\omega)$ under $C_3$ is given by\cite{supplementary_materials}
\begin{equation}
\label{eq:intensityequation}
I_\mathrm{PS}(2\omega) = (A_0 + A_1\cos{(3\phi)} + A_2\sin{(3\phi)})^2,
\end{equation}
where $A_0$, $A_1$, and $A_2$ are functions of the susceptibility elements $\chi^S_{ijk}$.

Symmetry considerations\cite{supplementary_materials} show that $A_0$ and $A_1$ vanish identically in the presence of mirror symmetry. The absence of these terms lead to the six-fold symmetry in the P$_{\mathrm{in}}$-S$_{\mathrm{out}}$ channel and its alignment with the crystallographic axes as seen in Fig.~\ref{fig:fig1}(a).
However, these terms can adopt nonzero values when mirror symmetry is broken.
Below $T_{\mathrm{IC-NC}}$, the RA-SHG intensity therefore exhibits a three-fold rather than six-fold symmetric pattern, arising from a nonzero $A_0$. 
The effect of a nonzero $A_1$ is to rotate the RA-SHG intensity away from the high-symmetry axes, but we observe this coefficient to be zero within the resolution of our instrument.
A negligible rotation of the SHG pattern should be expected, as the atomic positions of the Ta atoms only contract towards a central Ta atom and do not rotate away from their high symmetry axes to break the mirror symmetry (see Fig.~\ref{fig:fig0}(c)-(d)).

The three-fold nature of the RA-SHG intensity can be quantified experimentally by performing a spectral (sine) decomposition of the intensity and extracting the third Fourier coefficient, $\mathscr{I}_{PS}^{(3)}$, of $I_\mathrm{PS}(2\omega)$\cite{supplementary_materials}.
Figure~\ref{fig:fig1}(c) shows that $\mathscr{I}_{PS}^{(3)}$ appears discontinuously below $T_{\mathrm{IC-NC}}$, consistent with the first-order nature of the phase transition.
Taken together, the above considerations confirm that $\mathscr{I}_{PS}^{(3)}$ is a binary indicator of broken mirror symmetry in 1$T$-\ce{TaS2}.

Having established that RA-SHG is sensitive to the breaking of vertical mirror plane symmetry in 1$T$-\ce{TaS2}, we now seek to demonstrate that it can differentiate between CDW configurations of opposite planar chirality in the NC phase.
To do so, we generate two samples with opposite planar chirality by cleaving a single sample of 1$T$-\ce{TaS2}.
We then perform RA-SHG on both sides of the same cleave.
Referring to Figs.~\ref{fig:fig0}(c) and~\ref{fig:fig0}(d), cleaving the sample is equivalent to performing a $180^\circ$ rotation about the $x$-axis, which in a single layer is equivalent to a mirror reflection about $M_{xz}$.

Figure~\ref{fig:fig2} shows the results of RA-SHG measurements in the P$_\mathrm{in}$-P$_\mathrm{out}$ and P$_\mathrm{in}$-S$_\mathrm{out}$ polarization channels as functions of $\phi$, where the two mirror images are labeled $\alpha$ and $\beta$.
As shown in the figure, whether the CDW configuration was $\alpha$ or $\beta$ is indicated in RA-SHG by the orientation (up or down) of the pattern in the P$_\mathrm{in}$-S$_\mathrm{out}$ channel, which is determined by the sign of $A_0$ in Eq.~\ref{eq:intensityequation}\cite{supplementary_materials}.
This feature constitutes an experimental observable capable of identifying the sense associated with broken mirror symmetry in 1$T$-\ce{TaS2}. 

To validate the analysis contained above, we first fit the data for the $\alpha$ structure using the surface point group $C_{3}$ to generate a tensor $\chi^\alpha_{ijk}$ with numerical coefficients.
The fit is shown in Fig.~\ref{fig:fig2}(a).
Then, we transform $\chi^\alpha_{ijk}$ by a mirror reflection about $M_{xz}$ to generate $\chi_{ijk}^\beta$\cite{supplementary_materials}; with this transformation, we find that the rotational anisotropy simulated using $\chi_{ijk}^\beta$ collapses onto the measured signal, as shown in Fig.~\ref{fig:fig2}(b).

While the data contained in this work is specific to 1$T$-\ce{TaS2}, the analysis is applicable to a variety of different phase transitions involving broken mirror symmetry.
By understanding which Fourier coefficients of the SHG intensity adopt nonzero values in the low-symmetry phase, one can show that almost every structural phase transition involving broken mirror symmetry implies a measurable change in the symmetry of the SHG pattern (i.e. beyond a simple change in overall intensity).
We have performed a symmetry analysis of all possible phase transitions involving spontaneously broken mirror symmetry and in each case identified the relevant experimental indicator(s).
The results of this analysis can be found in the supplement\cite{supplementary_materials}.

\begin{figure}
\includegraphics{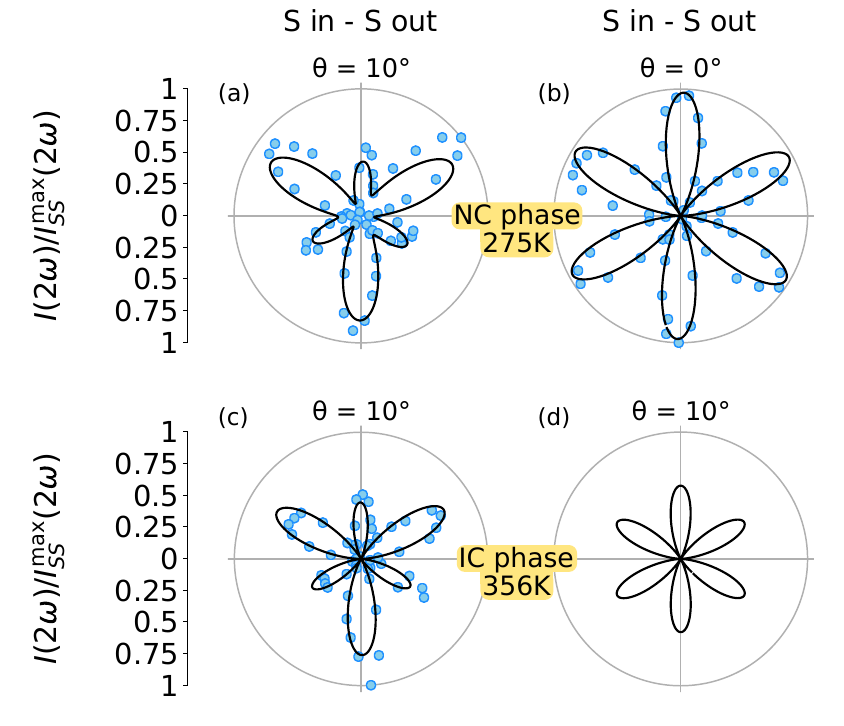}
\caption{\label{fig:fig3} Second harmonic intensity as a function of $\phi$ in the S$_\mathrm{in}$-S$_\mathrm{out}$ polarization channel.
Temperatures, phases and incident angles ($\theta$) are indicated on the figure.
Also shown are fitting curves using the point group assignments
(a) $C_3$ surface and $S_6$ bulk,
(b) $C_3$ surface and $S_6$ bulk,
(c) $C_{3v}$ surface and $S_6$ bulk.
(d) Best fit to data in (c) using $C_{3v}$ surface and $D_{3d}$ bulk.
Data in all plots was normalized to one for clariy.
Other polarization channels are depicted in the supplement\cite{supplementary_materials}.
%
}
\end{figure}

The final observation of this work concerns the contribution from the bulk of the sample to the measured RA-SHG signal.
As mentioned above, we find that in order to fit the data in the S$_\mathrm{in}$-S$_\mathrm{out}$ polarization channel correctly, we need to introduce an additional bulk quadrupole contribution to the SHG signal.
This contribution, which is described by the effective polarization $\nabla_j Q_{ij} = 2i\chi_{ijkl}^Qk_j E_k E_l$, is the next-lowest order contribution to SHG and is allowed in the presence of inversion symmetry~\cite{kumar_magnetic_2017, shen}.
Importantly, the quadrupole contribution is generated by the entire illumination volume and is therefore insensitive to the surface symmetry.
Quadrupole SHG can be identified by examining the $\theta$ dependence of the SHG intensity in the S$_\mathrm{in}$-S$_\mathrm{out}$ channel.
For purely electric dipole SHG, the symmetry in this channel does not depend on the incident angle.
The $\theta$ dependence depicted in Figs.~\ref{fig:fig3}(a) and~\ref{fig:fig3}(b) therefore establishes the presence of a quadrupole contribution in our signal.

According to diffraction measurements~\cite{fung_application_1980}, the correct surface and bulk point groups in the NC phase are $C_3$ and $S_6$, respectively.
Fig.~\ref{fig:fig3}(a) shows our results in this phase, which are consistent with this assignment.
To understand the three-fold symmetry in Fig.~\ref{fig:fig3}(a), we note that the $\phi$ dependence of $I_\mathrm{SS}(2\omega)$ in this symmetry assignment is given by\cite{supplementary_materials}
\begin{equation}
\label{eq:qintensityequation}
I_\mathrm{SS}(2\omega) = (B_0+B_1\cos{(3\phi)}+B_2\sin{(3\phi)})^2,
\end{equation}
where $B_1$ and $B_2$ are functions of the susceptibility elements $\chi^S_{ijk}$ and $\chi^Q_{ijkl}$, but $B_0$ depends on $\chi^Q_{ijkl}$ only (and is zero when the quadrupole contribution is ignored).
$B_0$ is then a probe of the bulk structure only and is not affected by the surface symmetry.
When mirror symmetry is broken, all three coefficients are allowed and the function is three-fold symmetric as in Fig.~\ref{fig:fig3}(a).

However, symmetry considerations\cite{supplementary_materials} show that $B_0$ is zero in the presence of mirror symmetry.
If bulk mirror symmetry were fully restored in the IC phase, we would therefore expect the signal to be six-fold symmetric.
Fig.~\ref{fig:fig3}(c) shows that the RA-SHG instead remains three-fold symmetric in this phase, suggesting that the bulk breaks mirror symmetry.
On the other hand, the highest-symmetry surface point group consistent with Figs.~\ref{fig:fig1}(a) and \ref{fig:fig3}(c) is $C_{3v}$, which retains mirror symmetry.
With RA-SHG, it is not possible to deduce the full crystal structure, but we speculate that this discrepancy between surface and bulk symmetries might be attributable to the particular stacking arrangement associated with the CDW above $T_{\mathrm{IC-NC}}$\cite{supplementary_materials}.
This would explain why the surface component, which is measuring the local structure at the surface of the sample~\cite{bloembergen_optical_1968}, is consistent with the existence of mirror symmetry, whereas the bulk component, which measures the global structure of many layers, is not.

In summary, we have demonstrated here that RA-SHG can be used to identify broken mirror symmetry in crystalline materials.
In addition, we have found that RA-SHG can differentiate between structural configurations related by mirror reflection.
By considering the different symmetry constraints on the S$_\mathrm{in}$-S$_\mathrm{out}$ channel, we have also shown that RA-SHG is sensitive to broken mirror symmetry in the bulk of 1$T$-\ce{TaS2}, and have found evidence that the IC phase of this material breaks mirror symmetry in the bulk.
Importantly, our analysis is generalizable beyond the specific case of 1$T$-\ce{TaS2}, and therefore opens up the possiblity for RA-SHG to detect mirror symmetry-broken phases and their domain structures in other candidate materials.

We would like to thank Adrian Po and Liang Fu for helpful discussions regarding this work.
This work was supported by the Gordon and Betty Moore Foundation’s EPiQS Initiative Grant GBMF4540 to NG (data analysis and manuscript writing) and Grant GBMF3848 to JGC (instrumentation), Shell through the MIT Energy Initiative (experimental setup and material development) and through DARPA DSO under DRINQS program grant number D18AC00014 (sample growth and data taking).
L.Y. acknowledges support by the STC Center for Integrated Quantum Materials, NSF Grant No. DMR-1231319 and by the Tsinghua Education Foundation.
B.G. gratefully acknowledges the German Academic Exchange Service (DAAD) for supporting his research stay with a fellowship.

\bibliography{NCtas2}

\begin{thebibliography}{33}%
\makeatletter
\providecommand \@ifxundefined [1]{%
 \@ifx{#1\undefined}
}%
\providecommand \@ifnum [1]{%
 \ifnum #1\expandafter \@firstoftwo
 \else \expandafter \@secondoftwo
 \fi
}%
\providecommand \@ifx [1]{%
 \ifx #1\expandafter \@firstoftwo
 \else \expandafter \@secondoftwo
 \fi
}%
\providecommand \natexlab [1]{#1}%
\providecommand \enquote  [1]{``#1''}%
\providecommand \bibnamefont  [1]{#1}%
\providecommand \bibfnamefont [1]{#1}%
\providecommand \citenamefont [1]{#1}%
\providecommand \href@noop [0]{\@secondoftwo}%
\providecommand \href [0]{\begingroup \@sanitize@url \@href}%
\providecommand \@href[1]{\@@startlink{#1}\@@href}%
\providecommand \@@href[1]{\endgroup#1\@@endlink}%
\providecommand \@sanitize@url [0]{\catcode `\\12\catcode `\$12\catcode
  `\&12\catcode `\#12\catcode `\^12\catcode `\_12\catcode `\%12\relax}%
\providecommand \@@startlink[1]{}%
\providecommand \@@endlink[0]{}%
\providecommand \url  [0]{\begingroup\@sanitize@url \@url }%
\providecommand \@url [1]{\endgroup\@href {#1}{\urlprefix }}%
\providecommand \urlprefix  [0]{URL }%
\providecommand \Eprint [0]{\href }%
\providecommand \doibase [0]{http://dx.doi.org/}%
\providecommand \selectlanguage [0]{\@gobble}%
\providecommand \bibinfo  [0]{\@secondoftwo}%
\providecommand \bibfield  [0]{\@secondoftwo}%
\providecommand \translation [1]{[#1]}%
\providecommand \BibitemOpen [0]{}%
\providecommand \bibitemStop [0]{}%
\providecommand \bibitemNoStop [0]{.\EOS\space}%
\providecommand \EOS [0]{\spacefactor3000\relax}%
\providecommand \BibitemShut  [1]{\csname bibitem#1\endcsname}%
\let\auto@bib@innerbib\@empty
\bibitem [{\citenamefont {Sethna}(2006)}]{sethna}%
  \BibitemOpen
  \bibfield  {author} {\bibinfo {author} {\bibfnamefont {J.}~\bibnamefont
  {Sethna}},\ }\href@noop {} {\emph {\bibinfo {title} {Statistical mechanics:
  entropy, order parameters, and complexity}}}\ (\bibinfo  {publisher} {Oxford
  University Press},\ \bibinfo {year} {2006})\BibitemShut {NoStop}%
\bibitem [{\citenamefont {Thouless}(1998)}]{thouless}%
  \BibitemOpen
  \bibfield  {author} {\bibinfo {author} {\bibfnamefont {D.}~\bibnamefont
  {Thouless}},\ }\href@noop {} {\emph {\bibinfo {title} {Topological quantum
  numbers in nonrelativistic physics}}}\ (\bibinfo  {publisher} {World
  Scientific},\ \bibinfo {year} {1998})\BibitemShut {NoStop}%
\bibitem [{\citenamefont {Belinicher}\ and\ \citenamefont
  {Sturman}(1980)}]{belinicher1980}%
  \BibitemOpen
  \bibfield  {author} {\bibinfo {author} {\bibfnamefont {V.}~\bibnamefont
  {Belinicher}}\ and\ \bibinfo {author} {\bibfnamefont {B.}~\bibnamefont
  {Sturman}},\ }\href@noop {} {\bibfield  {journal} {\bibinfo  {journal}
  {Soviet Physics Uspekhi}\ }\textbf {\bibinfo {volume} {23}},\ \bibinfo
  {pages} {199} (\bibinfo {year} {1980})}\BibitemShut {NoStop}%
\bibitem [{\citenamefont {Teo}\ \emph {et~al.}(2008)\citenamefont {Teo},
  \citenamefont {Fu},\ and\ \citenamefont {Kane}}]{teo2008}%
  \BibitemOpen
  \bibfield  {author} {\bibinfo {author} {\bibfnamefont {J.~C.~Y.}\
  \bibnamefont {Teo}}, \bibinfo {author} {\bibfnamefont {L.}~\bibnamefont
  {Fu}}, \ and\ \bibinfo {author} {\bibfnamefont {C.~L.}\ \bibnamefont
  {Kane}},\ }\href {\doibase 10.1103/PhysRevB.78.045426} {\bibfield  {journal}
  {\bibinfo  {journal} {Phys. Rev. B}\ }\textbf {\bibinfo {volume} {78}},\
  \bibinfo {pages} {045426} (\bibinfo {year} {2008})}\BibitemShut {NoStop}%
\bibitem [{\citenamefont {Hsieh}\ \emph {et~al.}(2012)\citenamefont {Hsieh},
  \citenamefont {Lin}, \citenamefont {Liu}, \citenamefont {Duan}, \citenamefont
  {Bansil},\ and\ \citenamefont {Fu}}]{hsieh_topological_2012}%
  \BibitemOpen
  \bibfield  {author} {\bibinfo {author} {\bibfnamefont {T.~H.}\ \bibnamefont
  {Hsieh}}, \bibinfo {author} {\bibfnamefont {H.}~\bibnamefont {Lin}}, \bibinfo
  {author} {\bibfnamefont {J.}~\bibnamefont {Liu}}, \bibinfo {author}
  {\bibfnamefont {W.}~\bibnamefont {Duan}}, \bibinfo {author} {\bibfnamefont
  {A.}~\bibnamefont {Bansil}}, \ and\ \bibinfo {author} {\bibfnamefont
  {L.}~\bibnamefont {Fu}},\ }\href {\doibase 10.1038/ncomms1969} {\bibfield
  {journal} {\bibinfo  {journal} {Nature Communications}\ }\textbf {\bibinfo
  {volume} {3}} (\bibinfo {year} {2012}),\ 10.1038/ncomms1969}\BibitemShut
  {NoStop}%
\bibitem [{\citenamefont {Fu}(2011)}]{fu_topological_2011}%
  \BibitemOpen
  \bibfield  {author} {\bibinfo {author} {\bibfnamefont {L.}~\bibnamefont
  {Fu}},\ }\href {\doibase 10.1103/PhysRevLett.106.106802} {\bibfield
  {journal} {\bibinfo  {journal} {Physical Review Letters}\ }\textbf {\bibinfo
  {volume} {106}} (\bibinfo {year} {2011}),\
  10.1103/PhysRevLett.106.106802}\BibitemShut {NoStop}%
\bibitem [{\citenamefont {Hlobil}\ \emph {et~al.}(2015)\citenamefont {Hlobil},
  \citenamefont {Maharaj}, \citenamefont {Hosur}, \citenamefont {Shapiro},
  \citenamefont {Fisher},\ and\ \citenamefont
  {Raghu}}]{hlobil_elastoconductivity_2015}%
  \BibitemOpen
  \bibfield  {author} {\bibinfo {author} {\bibfnamefont {P.}~\bibnamefont
  {Hlobil}}, \bibinfo {author} {\bibfnamefont {A.~V.}\ \bibnamefont {Maharaj}},
  \bibinfo {author} {\bibfnamefont {P.}~\bibnamefont {Hosur}}, \bibinfo
  {author} {\bibfnamefont {M.~C.}\ \bibnamefont {Shapiro}}, \bibinfo {author}
  {\bibfnamefont {I.~R.}\ \bibnamefont {Fisher}}, \ and\ \bibinfo {author}
  {\bibfnamefont {S.}~\bibnamefont {Raghu}},\ }\href {\doibase
  10.1103/PhysRevB.92.035148} {\bibfield  {journal} {\bibinfo  {journal}
  {Physical Review B}\ }\textbf {\bibinfo {volume} {92}} (\bibinfo {year}
  {2015}),\ 10.1103/PhysRevB.92.035148}\BibitemShut {NoStop}%
\bibitem [{\citenamefont {Leisure}\ and\ \citenamefont
  {Willis}(1997)}]{leisure1997}%
  \BibitemOpen
  \bibfield  {author} {\bibinfo {author} {\bibfnamefont {R.~G.}\ \bibnamefont
  {Leisure}}\ and\ \bibinfo {author} {\bibfnamefont {F.}~\bibnamefont
  {Willis}},\ }\href@noop {} {\bibfield  {journal} {\bibinfo  {journal}
  {Journal of Physics: Condensed Matter}\ }\textbf {\bibinfo {volume} {9}},\
  \bibinfo {pages} {6001} (\bibinfo {year} {1997})}\BibitemShut {NoStop}%
\bibitem [{\citenamefont {Migliori}\ \emph {et~al.}(1993)\citenamefont
  {Migliori}, \citenamefont {Sarrao}, \citenamefont {Visscher}, \citenamefont
  {Bell}, \citenamefont {Lei}, \citenamefont {Fisk},\ and\ \citenamefont
  {Leisure}}]{migliori1993}%
  \BibitemOpen
  \bibfield  {author} {\bibinfo {author} {\bibfnamefont {A.}~\bibnamefont
  {Migliori}}, \bibinfo {author} {\bibfnamefont {J.}~\bibnamefont {Sarrao}},
  \bibinfo {author} {\bibfnamefont {W.~M.}\ \bibnamefont {Visscher}}, \bibinfo
  {author} {\bibfnamefont {T.}~\bibnamefont {Bell}}, \bibinfo {author}
  {\bibfnamefont {M.}~\bibnamefont {Lei}}, \bibinfo {author} {\bibfnamefont
  {Z.}~\bibnamefont {Fisk}}, \ and\ \bibinfo {author} {\bibfnamefont {R.~G.}\
  \bibnamefont {Leisure}},\ }\href@noop {} {\bibfield  {journal} {\bibinfo
  {journal} {Physica B: Condensed Matter}\ }\textbf {\bibinfo {volume} {183}},\
  \bibinfo {pages} {1} (\bibinfo {year} {1993})}\BibitemShut {NoStop}%
\bibitem [{\citenamefont {Bacon}(1966)}]{bacon1966}%
  \BibitemOpen
  \bibfield  {author} {\bibinfo {author} {\bibfnamefont {G.~E.}\ \bibnamefont
  {Bacon}},\ }\href@noop {} {\emph {\bibinfo {title} {X-ray and Neutron
  Diffraction}}}\ (\bibinfo  {publisher} {Elsevier},\ \bibinfo {year}
  {1966})\BibitemShut {NoStop}%
\bibitem [{\citenamefont {Dorset}(2013)}]{dorset2013}%
  \BibitemOpen
  \bibfield  {author} {\bibinfo {author} {\bibfnamefont {D.~L.}\ \bibnamefont
  {Dorset}},\ }\href@noop {} {\emph {\bibinfo {title} {Structural electron
  crystallography}}}\ (\bibinfo  {publisher} {Springer Science \& Business
  Media},\ \bibinfo {year} {2013})\BibitemShut {NoStop}%
\bibitem [{\citenamefont {Heinz}\ \emph {et~al.}(1985)\citenamefont {Heinz},
  \citenamefont {Loy},\ and\ \citenamefont {Thompson}}]{heinz_study_1985}%
  \BibitemOpen
  \bibfield  {author} {\bibinfo {author} {\bibfnamefont {T.~F.}\ \bibnamefont
  {Heinz}}, \bibinfo {author} {\bibfnamefont {M.~M.~T.}\ \bibnamefont {Loy}}, \
  and\ \bibinfo {author} {\bibfnamefont {W.~A.}\ \bibnamefont {Thompson}},\
  }\href {\doibase 10.1103/PhysRevLett.54.63} {\bibfield  {journal} {\bibinfo
  {journal} {Phys. Rev. Lett.}\ }\textbf {\bibinfo {volume} {54}},\ \bibinfo
  {pages} {63} (\bibinfo {year} {1985})}\BibitemShut {NoStop}%
\bibitem [{\citenamefont {Torchinsky}\ and\ \citenamefont
  {Hsieh}(2017)}]{kumar_magnetic_2017}%
  \BibitemOpen
  \bibfield  {author} {\bibinfo {author} {\bibfnamefont {D.~H.}\ \bibnamefont
  {Torchinsky}}\ and\ \bibinfo {author} {\bibfnamefont {D.}~\bibnamefont
  {Hsieh}},\ }in\ \href {\doibase 10.1007/978-3-662-52780-1} {\emph {\bibinfo
  {booktitle} {Magnetic {Characterization} {Techniques} for
  {Nanomaterials}}}},\ \bibinfo {editor} {edited by\ \bibinfo {editor}
  {\bibfnamefont {C.~S.}\ \bibnamefont {Kumar}}}\ (\bibinfo  {publisher}
  {Springer Berlin Heidelberg},\ \bibinfo {address} {Berlin, Heidelberg},\
  \bibinfo {year} {2017})\ pp.\ \bibinfo {pages} {1--40}\BibitemShut {NoStop}%
\bibitem [{\citenamefont {Harter}\ \emph {et~al.}(2017)\citenamefont {Harter},
  \citenamefont {Zhao}, \citenamefont {Yan}, \citenamefont {Mandrus},\ and\
  \citenamefont {Hsieh}}]{harter_parity-breaking_2017}%
  \BibitemOpen
  \bibfield  {author} {\bibinfo {author} {\bibfnamefont {J.~W.}\ \bibnamefont
  {Harter}}, \bibinfo {author} {\bibfnamefont {Z.~Y.}\ \bibnamefont {Zhao}},
  \bibinfo {author} {\bibfnamefont {J.-Q.}\ \bibnamefont {Yan}}, \bibinfo
  {author} {\bibfnamefont {D.~G.}\ \bibnamefont {Mandrus}}, \ and\ \bibinfo
  {author} {\bibfnamefont {D.}~\bibnamefont {Hsieh}},\ }\href {\doibase
  10.1126/science.aad1188} {\bibfield  {journal} {\bibinfo  {journal}
  {Science}\ }\textbf {\bibinfo {volume} {356}},\ \bibinfo {pages} {295}
  (\bibinfo {year} {2017})}\BibitemShut {NoStop}%
\bibitem [{\citenamefont {Wilson}\ \emph {et~al.}(1975)\citenamefont {Wilson},
  \citenamefont {Di~Salvo},\ and\ \citenamefont
  {Mahajan}}]{wilson_charge-density_1975}%
  \BibitemOpen
  \bibfield  {author} {\bibinfo {author} {\bibfnamefont {J.}~\bibnamefont
  {Wilson}}, \bibinfo {author} {\bibfnamefont {F.}~\bibnamefont {Di~Salvo}}, \
  and\ \bibinfo {author} {\bibfnamefont {S.}~\bibnamefont {Mahajan}},\ }\href
  {\doibase 10.1080/00018737500101391} {\bibfield  {journal} {\bibinfo
  {journal} {Advances in Physics}\ }\textbf {\bibinfo {volume} {24}},\ \bibinfo
  {pages} {117} (\bibinfo {year} {1975})}\BibitemShut {NoStop}%
\bibitem [{\citenamefont {Zong}\ \emph {et~al.}(2018)\citenamefont {Zong},
  \citenamefont {Shen}, \citenamefont {Kogar}, \citenamefont {Ye},
  \citenamefont {Marks}, \citenamefont {Chowdhury}, \citenamefont {Rohwer},
  \citenamefont {Freelon}, \citenamefont {Weathersby}, \citenamefont {Li},
  \citenamefont {Yang}, \citenamefont {Checkelsky}, \citenamefont {Wang},\ and\
  \citenamefont {Gedik}}]{zong_ultrafast_2018}%
  \BibitemOpen
  \bibfield  {author} {\bibinfo {author} {\bibfnamefont {A.}~\bibnamefont
  {Zong}}, \bibinfo {author} {\bibfnamefont {X.}~\bibnamefont {Shen}}, \bibinfo
  {author} {\bibfnamefont {A.}~\bibnamefont {Kogar}}, \bibinfo {author}
  {\bibfnamefont {L.}~\bibnamefont {Ye}}, \bibinfo {author} {\bibfnamefont
  {C.}~\bibnamefont {Marks}}, \bibinfo {author} {\bibfnamefont
  {D.}~\bibnamefont {Chowdhury}}, \bibinfo {author} {\bibfnamefont
  {T.}~\bibnamefont {Rohwer}}, \bibinfo {author} {\bibfnamefont
  {B.}~\bibnamefont {Freelon}}, \bibinfo {author} {\bibfnamefont
  {S.}~\bibnamefont {Weathersby}}, \bibinfo {author} {\bibfnamefont
  {R.}~\bibnamefont {Li}}, \bibinfo {author} {\bibfnamefont {J.}~\bibnamefont
  {Yang}}, \bibinfo {author} {\bibfnamefont {J.}~\bibnamefont {Checkelsky}},
  \bibinfo {author} {\bibfnamefont {X.}~\bibnamefont {Wang}}, \ and\ \bibinfo
  {author} {\bibfnamefont {N.}~\bibnamefont {Gedik}},\ }\href {\doibase
  10.1126/sciadv.aau5501} {\bibfield  {journal} {\bibinfo  {journal} {Science
  Advances}\ }\textbf {\bibinfo {volume} {4}} (\bibinfo {year} {2018}),\
  10.1126/sciadv.aau5501}\BibitemShut {NoStop}%
\bibitem [{\citenamefont {Spijkerman}\ \emph {et~al.}(1997)\citenamefont
  {Spijkerman}, \citenamefont {de~Boer}, \citenamefont {Meetsma}, \citenamefont
  {Wiegers},\ and\ \citenamefont {van Smaalen}}]{spijkerman_x-ray_1997}%
  \BibitemOpen
  \bibfield  {author} {\bibinfo {author} {\bibfnamefont {A.}~\bibnamefont
  {Spijkerman}}, \bibinfo {author} {\bibfnamefont {J.~L.}\ \bibnamefont
  {de~Boer}}, \bibinfo {author} {\bibfnamefont {A.}~\bibnamefont {Meetsma}},
  \bibinfo {author} {\bibfnamefont {G.~A.}\ \bibnamefont {Wiegers}}, \ and\
  \bibinfo {author} {\bibfnamefont {S.}~\bibnamefont {van Smaalen}},\ }\href
  {\doibase 10.1103/PhysRevB.56.13757} {\bibfield  {journal} {\bibinfo
  {journal} {Physical Review B}\ }\textbf {\bibinfo {volume} {56}},\ \bibinfo
  {pages} {13757} (\bibinfo {year} {1997})}\BibitemShut {NoStop}%
\bibitem [{\citenamefont {Scruby}\ \emph {et~al.}(1975)\citenamefont {Scruby},
  \citenamefont {Williams},\ and\ \citenamefont {Parry}}]{scruby_role_1975}%
  \BibitemOpen
  \bibfield  {author} {\bibinfo {author} {\bibfnamefont {C.~B.}\ \bibnamefont
  {Scruby}}, \bibinfo {author} {\bibfnamefont {P.~M.}\ \bibnamefont
  {Williams}}, \ and\ \bibinfo {author} {\bibfnamefont {G.~S.}\ \bibnamefont
  {Parry}},\ }\href {\doibase 10.1080/14786437508228930} {\bibfield  {journal}
  {\bibinfo  {journal} {Philosophical Magazine}\ }\textbf {\bibinfo {volume}
  {31}},\ \bibinfo {pages} {255} (\bibinfo {year} {1975})}\BibitemShut
  {NoStop}%
\bibitem [{\citenamefont {Fung}\ \emph {et~al.}(1980)\citenamefont {Fung},
  \citenamefont {Steeds},\ and\ \citenamefont {Eades}}]{fung_application_1980}%
  \BibitemOpen
  \bibfield  {author} {\bibinfo {author} {\bibfnamefont {K.}~\bibnamefont
  {Fung}}, \bibinfo {author} {\bibfnamefont {J.}~\bibnamefont {Steeds}}, \ and\
  \bibinfo {author} {\bibfnamefont {J.}~\bibnamefont {Eades}},\ }\href
  {\doibase 10.1016/0378-4363(80)90208-9} {\bibfield  {journal} {\bibinfo
  {journal} {Physica B+C}\ }\textbf {\bibinfo {volume} {99}},\ \bibinfo {pages}
  {47} (\bibinfo {year} {1980})}\BibitemShut {NoStop}%
\bibitem [{\citenamefont {Wu}\ and\ \citenamefont
  {Lieber}(1989)}]{wu_hexagonal_1989}%
  \BibitemOpen
  \bibfield  {author} {\bibinfo {author} {\bibfnamefont {X.~L.}\ \bibnamefont
  {Wu}}\ and\ \bibinfo {author} {\bibfnamefont {C.~M.}\ \bibnamefont
  {Lieber}},\ }\href {\doibase 10.1126/science.243.4899.1703} {\bibfield
  {journal} {\bibinfo  {journal} {Science}\ }\textbf {\bibinfo {volume}
  {243}},\ \bibinfo {pages} {1703} (\bibinfo {year} {1989})}\BibitemShut
  {NoStop}%
\bibitem [{\citenamefont {Ishiguro}\ and\ \citenamefont
  {Sato}(1995)}]{ishiguro_high-resolution_1995}%
  \BibitemOpen
  \bibfield  {author} {\bibinfo {author} {\bibfnamefont {T.}~\bibnamefont
  {Ishiguro}}\ and\ \bibinfo {author} {\bibfnamefont {H.}~\bibnamefont
  {Sato}},\ }\href {\doibase 10.1103/PhysRevB.52.759} {\bibfield  {journal}
  {\bibinfo  {journal} {Physical Review B}\ }\textbf {\bibinfo {volume} {52}},\
  \bibinfo {pages} {759} (\bibinfo {year} {1995})}\BibitemShut {NoStop}%
\bibitem [{sup()}]{supplementary_materials}%
  \BibitemOpen
  \href@noop {} {}\bibinfo {note} {See Supplemental Material at [URL inserted
  by publisher] for further details.}\BibitemShut {Stop}%
\bibitem [{\citenamefont {Bovet}\ \emph {et~al.}(2004)\citenamefont {Bovet},
  \citenamefont {Popović}, \citenamefont {Clerc}, \citenamefont {Koitzsch},
  \citenamefont {Probst}, \citenamefont {Bucher}, \citenamefont {Berger},
  \citenamefont {Naumović},\ and\ \citenamefont
  {Aebi}}]{bovet_pseudogapped_2004}%
  \BibitemOpen
  \bibfield  {author} {\bibinfo {author} {\bibfnamefont {M.}~\bibnamefont
  {Bovet}}, \bibinfo {author} {\bibfnamefont {D.}~\bibnamefont {Popović}},
  \bibinfo {author} {\bibfnamefont {F.}~\bibnamefont {Clerc}}, \bibinfo
  {author} {\bibfnamefont {C.}~\bibnamefont {Koitzsch}}, \bibinfo {author}
  {\bibfnamefont {U.}~\bibnamefont {Probst}}, \bibinfo {author} {\bibfnamefont
  {E.}~\bibnamefont {Bucher}}, \bibinfo {author} {\bibfnamefont
  {H.}~\bibnamefont {Berger}}, \bibinfo {author} {\bibfnamefont
  {D.}~\bibnamefont {Naumović}}, \ and\ \bibinfo {author} {\bibfnamefont
  {P.}~\bibnamefont {Aebi}},\ }\href {\doibase 10.1103/PhysRevB.69.125117}
  {\bibfield  {journal} {\bibinfo  {journal} {Physical Review B}\ }\textbf
  {\bibinfo {volume} {69}} (\bibinfo {year} {2004}),\
  10.1103/PhysRevB.69.125117}\BibitemShut {NoStop}%
\bibitem [{\citenamefont {H.~Shiba}(1986)}]{shiba_phenomenological_1986}%
  \BibitemOpen
  \bibfield  {author} {\bibinfo {author} {\bibfnamefont {K.~N.}\ \bibnamefont
  {H.~Shiba}},\ }in\ \href@noop {} {\emph {\bibinfo {booktitle} {Structural
  {P}hase {T}ransitions in {L}ayered {T}ransition {M}etal {C}ompounds}}},\
  \bibinfo {editor} {edited by\ \bibinfo {editor} {\bibfnamefont
  {K.}~\bibnamefont {Motizuki}}}\ (\bibinfo  {publisher} {D. Reidel Publishing
  Company},\ \bibinfo {year} {1986})\ pp.\ \bibinfo {pages}
  {175--266}\BibitemShut {NoStop}%
\bibitem [{\citenamefont {Harter}\ \emph {et~al.}(2015)\citenamefont {Harter},
  \citenamefont {Niu}, \citenamefont {Woss},\ and\ \citenamefont
  {Hsieh}}]{harter_high-speed_2015}%
  \BibitemOpen
  \bibfield  {author} {\bibinfo {author} {\bibfnamefont {J.~W.}\ \bibnamefont
  {Harter}}, \bibinfo {author} {\bibfnamefont {L.}~\bibnamefont {Niu}},
  \bibinfo {author} {\bibfnamefont {A.~J.}\ \bibnamefont {Woss}}, \ and\
  \bibinfo {author} {\bibfnamefont {D.}~\bibnamefont {Hsieh}},\ }\href
  {\doibase 10.1364/OL.40.004671} {\bibfield  {journal} {\bibinfo  {journal}
  {Optics Letters}\ }\textbf {\bibinfo {volume} {40}},\ \bibinfo {pages} {4671}
  (\bibinfo {year} {2015})}\BibitemShut {NoStop}%
\bibitem [{\citenamefont {Lu}\ \emph {et~al.}(2019)\citenamefont {Lu},
  \citenamefont {Tran},\ and\ \citenamefont {Torchinsky}}]{lu_fast_2018}%
  \BibitemOpen
  \bibfield  {author} {\bibinfo {author} {\bibfnamefont {B.}~\bibnamefont
  {Lu}}, \bibinfo {author} {\bibfnamefont {J.~D.}\ \bibnamefont {Tran}}, \ and\
  \bibinfo {author} {\bibfnamefont {D.~H.}\ \bibnamefont {Torchinsky}},\ }\href
  {\doibase 10.1063/1.5080965} {\bibfield  {journal} {\bibinfo  {journal}
  {Review of Scientific Instruments}\ }\textbf {\bibinfo {volume} {90}},\
  \bibinfo {pages} {053102} (\bibinfo {year} {2019})},\ \Eprint
  {http://arxiv.org/abs/https://doi.org/10.1063/1.5080965}
  {https://doi.org/10.1063/1.5080965} \BibitemShut {NoStop}%
\bibitem [{\citenamefont {Torchinsky}\ \emph {et~al.}(2015)\citenamefont
  {Torchinsky}, \citenamefont {Chu}, \citenamefont {Zhao}, \citenamefont
  {Perkins}, \citenamefont {Sizyuk}, \citenamefont {Qi}, \citenamefont {Cao},\
  and\ \citenamefont {Hsieh}}]{torchinsky_structural_2015}%
  \BibitemOpen
  \bibfield  {author} {\bibinfo {author} {\bibfnamefont {D.~H.}\ \bibnamefont
  {Torchinsky}}, \bibinfo {author} {\bibfnamefont {H.}~\bibnamefont {Chu}},
  \bibinfo {author} {\bibfnamefont {L.}~\bibnamefont {Zhao}}, \bibinfo {author}
  {\bibfnamefont {N.~B.}\ \bibnamefont {Perkins}}, \bibinfo {author}
  {\bibfnamefont {Y.}~\bibnamefont {Sizyuk}}, \bibinfo {author} {\bibfnamefont
  {T.}~\bibnamefont {Qi}}, \bibinfo {author} {\bibfnamefont {G.}~\bibnamefont
  {Cao}}, \ and\ \bibinfo {author} {\bibfnamefont {D.}~\bibnamefont {Hsieh}},\
  }\href {\doibase 10.1103/PhysRevLett.114.096404} {\bibfield  {journal}
  {\bibinfo  {journal} {Physical Review Letters}\ }\textbf {\bibinfo {volume}
  {114}} (\bibinfo {year} {2015}),\ 10.1103/PhysRevLett.114.096404}\BibitemShut
  {NoStop}%
\bibitem [{\citenamefont {Lu}\ and\ \citenamefont
  {Torchinsky}(2018)}]{lu_fourier_2018}%
  \BibitemOpen
  \bibfield  {author} {\bibinfo {author} {\bibfnamefont {B.}~\bibnamefont
  {Lu}}\ and\ \bibinfo {author} {\bibfnamefont {D.~H.}\ \bibnamefont
  {Torchinsky}},\ }\href {\doibase 10.1364/OE.26.033192} {\bibfield  {journal}
  {\bibinfo  {journal} {Opt. Express}\ }\textbf {\bibinfo {volume} {26}},\
  \bibinfo {pages} {33192} (\bibinfo {year} {2018})}\BibitemShut {NoStop}%
\bibitem [{\citenamefont {Boyd}(2008)}]{boyd}%
  \BibitemOpen
  \bibfield  {author} {\bibinfo {author} {\bibfnamefont {R.}~\bibnamefont
  {Boyd}},\ }\href@noop {} {\emph {\bibinfo {title} {Nonlinear Optics}}},\
  \bibinfo {edition} {3rd}\ ed.\ (\bibinfo  {publisher} {Academic Press},\
  \bibinfo {year} {2008})\BibitemShut {NoStop}%
\bibitem [{\citenamefont {Powell}(2010)}]{powell}%
  \BibitemOpen
  \bibfield  {author} {\bibinfo {author} {\bibfnamefont {R.~C.}\ \bibnamefont
  {Powell}},\ }\href@noop {} {\emph {\bibinfo {title} {Symmetry, Group Theory,
  and the Physical Properties of Crystals}}},\ \bibinfo {edition} {1st}\ ed.,\
  \bibinfo {series} {Lecture Notes in Physics}, Vol.\ \bibinfo {volume} {824}\
  (\bibinfo  {publisher} {Springer-Verlag},\ \bibinfo {address} {New York},\
  \bibinfo {year} {2010})\BibitemShut {NoStop}%
\bibitem [{\citenamefont {Bloembergen}\ \emph {et~al.}(1968)\citenamefont
  {Bloembergen}, \citenamefont {Chang}, \citenamefont {Jha},\ and\
  \citenamefont {Lee}}]{bloembergen_optical_1968}%
  \BibitemOpen
  \bibfield  {author} {\bibinfo {author} {\bibfnamefont {N.}~\bibnamefont
  {Bloembergen}}, \bibinfo {author} {\bibfnamefont {R.~K.}\ \bibnamefont
  {Chang}}, \bibinfo {author} {\bibfnamefont {S.~S.}\ \bibnamefont {Jha}}, \
  and\ \bibinfo {author} {\bibfnamefont {C.~H.}\ \bibnamefont {Lee}},\ }\href
  {\doibase 10.1103/PhysRev.174.813} {\bibfield  {journal} {\bibinfo  {journal}
  {Physical Review}\ }\textbf {\bibinfo {volume} {174}},\ \bibinfo {pages}
  {813} (\bibinfo {year} {1968})}\BibitemShut {NoStop}%
\bibitem [{\citenamefont {Shen}(1984)}]{shen}%
  \BibitemOpen
  \bibfield  {author} {\bibinfo {author} {\bibfnamefont {Y.}~\bibnamefont
  {Shen}},\ }\href {https://books.google.com/books?id=qYIpAQAAMAAJ} {\emph
  {\bibinfo {title} {The Principles of Nonlinear Optics}}},\ Pure \& Applied
  Optics Series: 1-349\ (\bibinfo  {publisher} {Wiley},\ \bibinfo {year}
  {1984})\BibitemShut {NoStop}%
\bibitem [{\citenamefont {Torchinsky}\ \emph {et~al.}(2014)\citenamefont
  {Torchinsky}, \citenamefont {Chu}, \citenamefont {Qi}, \citenamefont {Cao},\
  and\ \citenamefont {Hsieh}}]{torchinsky_low_2014}%
  \BibitemOpen
  \bibfield  {author} {\bibinfo {author} {\bibfnamefont {D.~H.}\ \bibnamefont
  {Torchinsky}}, \bibinfo {author} {\bibfnamefont {H.}~\bibnamefont {Chu}},
  \bibinfo {author} {\bibfnamefont {T.}~\bibnamefont {Qi}}, \bibinfo {author}
  {\bibfnamefont {G.}~\bibnamefont {Cao}}, \ and\ \bibinfo {author}
  {\bibfnamefont {D.}~\bibnamefont {Hsieh}},\ }\href {\doibase
  10.1063/1.4891417} {\bibfield  {journal} {\bibinfo  {journal} {Review of
  Scientific Instruments}\ }\textbf {\bibinfo {volume} {85}},\ \bibinfo {pages}
  {083102} (\bibinfo {year} {2014})}\BibitemShut {NoStop}%
\end{thebibliography}%


\begin{thebibliography}{12}%
\makeatletter
\providecommand \@ifxundefined [1]{%
 \@ifx{#1\undefined}
}%
\providecommand \@ifnum [1]{%
 \ifnum #1\expandafter \@firstoftwo
 \else \expandafter \@secondoftwo
 \fi
}%
\providecommand \@ifx [1]{%
 \ifx #1\expandafter \@firstoftwo
 \else \expandafter \@secondoftwo
 \fi
}%
\providecommand \natexlab [1]{#1}%
\providecommand \enquote  [1]{``#1''}%
\providecommand \bibnamefont  [1]{#1}%
\providecommand \bibfnamefont [1]{#1}%
\providecommand \citenamefont [1]{#1}%
\providecommand \href@noop [0]{\@secondoftwo}%
\providecommand \href [0]{\begingroup \@sanitize@url \@href}%
\providecommand \@href[1]{\@@startlink{#1}\@@href}%
\providecommand \@@href[1]{\endgroup#1\@@endlink}%
\providecommand \@sanitize@url [0]{\catcode `\\12\catcode `\$12\catcode
  `\&12\catcode `\#12\catcode `\^12\catcode `\_12\catcode `\%12\relax}%
\providecommand \@@startlink[1]{}%
\providecommand \@@endlink[0]{}%
\providecommand \url  [0]{\begingroup\@sanitize@url \@url }%
\providecommand \@url [1]{\endgroup\@href {#1}{\urlprefix }}%
\providecommand \urlprefix  [0]{URL }%
\providecommand \Eprint [0]{\href }%
\providecommand \doibase [0]{http://dx.doi.org/}%
\providecommand \selectlanguage [0]{\@gobble}%
\providecommand \bibinfo  [0]{\@secondoftwo}%
\providecommand \bibfield  [0]{\@secondoftwo}%
\providecommand \translation [1]{[#1]}%
\providecommand \BibitemOpen [0]{}%
\providecommand \bibitemStop [0]{}%
\providecommand \bibitemNoStop [0]{.\EOS\space}%
\providecommand \EOS [0]{\spacefactor3000\relax}%
\providecommand \BibitemShut  [1]{\csname bibitem#1\endcsname}%
\let\auto@bib@innerbib\@empty
\bibitem [{\citenamefont {Boyd}(2008)}]{boyd}%
  \BibitemOpen
  \bibfield  {author} {\bibinfo {author} {\bibfnamefont {R.}~\bibnamefont
  {Boyd}},\ }\href@noop {} {\emph {\bibinfo {title} {Nonlinear Optics}}},\
  \bibinfo {edition} {3rd}\ ed.\ (\bibinfo  {publisher} {Academic Press},\
  \bibinfo {year} {2008})\BibitemShut {NoStop}%
\bibitem [{\citenamefont {Zong}\ \emph {et~al.}(2018)\citenamefont {Zong},
  \citenamefont {Shen}, \citenamefont {Kogar}, \citenamefont {Ye},
  \citenamefont {Marks}, \citenamefont {Chowdhury}, \citenamefont {Rohwer},
  \citenamefont {Freelon}, \citenamefont {Weathersby}, \citenamefont {Li},
  \citenamefont {Yang}, \citenamefont {Checkelsky}, \citenamefont {Wang},\ and\
  \citenamefont {Gedik}}]{zong_ultrafast_2018}%
  \BibitemOpen
  \bibfield  {author} {\bibinfo {author} {\bibfnamefont {A.}~\bibnamefont
  {Zong}}, \bibinfo {author} {\bibfnamefont {X.}~\bibnamefont {Shen}}, \bibinfo
  {author} {\bibfnamefont {A.}~\bibnamefont {Kogar}}, \bibinfo {author}
  {\bibfnamefont {L.}~\bibnamefont {Ye}}, \bibinfo {author} {\bibfnamefont
  {C.}~\bibnamefont {Marks}}, \bibinfo {author} {\bibfnamefont
  {D.}~\bibnamefont {Chowdhury}}, \bibinfo {author} {\bibfnamefont
  {T.}~\bibnamefont {Rohwer}}, \bibinfo {author} {\bibfnamefont
  {B.}~\bibnamefont {Freelon}}, \bibinfo {author} {\bibfnamefont
  {S.}~\bibnamefont {Weathersby}}, \bibinfo {author} {\bibfnamefont
  {R.}~\bibnamefont {Li}}, \bibinfo {author} {\bibfnamefont {J.}~\bibnamefont
  {Yang}}, \bibinfo {author} {\bibfnamefont {J.}~\bibnamefont {Checkelsky}},
  \bibinfo {author} {\bibfnamefont {X.}~\bibnamefont {Wang}}, \ and\ \bibinfo
  {author} {\bibfnamefont {N.}~\bibnamefont {Gedik}},\ }\href {\doibase
  10.1126/sciadv.aau5501} {\bibfield  {journal} {\bibinfo  {journal} {Science
  Advances}\ }\textbf {\bibinfo {volume} {4}} (\bibinfo {year} {2018}),\
  10.1126/sciadv.aau5501}\BibitemShut {NoStop}%
\bibitem [{\citenamefont {Wilson}\ \emph {et~al.}(1975)\citenamefont {Wilson},
  \citenamefont {Di~Salvo},\ and\ \citenamefont
  {Mahajan}}]{wilson_charge-density_1975}%
  \BibitemOpen
  \bibfield  {author} {\bibinfo {author} {\bibfnamefont {J.}~\bibnamefont
  {Wilson}}, \bibinfo {author} {\bibfnamefont {F.}~\bibnamefont {Di~Salvo}}, \
  and\ \bibinfo {author} {\bibfnamefont {S.}~\bibnamefont {Mahajan}},\ }\href
  {\doibase 10.1080/00018737500101391} {\bibfield  {journal} {\bibinfo
  {journal} {Advances in Physics}\ }\textbf {\bibinfo {volume} {24}},\ \bibinfo
  {pages} {117} (\bibinfo {year} {1975})}\BibitemShut {NoStop}%
\bibitem [{\citenamefont {Bovet}\ \emph {et~al.}(2004)\citenamefont {Bovet},
  \citenamefont {Popović}, \citenamefont {Clerc}, \citenamefont {Koitzsch},
  \citenamefont {Probst}, \citenamefont {Bucher}, \citenamefont {Berger},
  \citenamefont {Naumović},\ and\ \citenamefont
  {Aebi}}]{bovet_pseudogapped_2004}%
  \BibitemOpen
  \bibfield  {author} {\bibinfo {author} {\bibfnamefont {M.}~\bibnamefont
  {Bovet}}, \bibinfo {author} {\bibfnamefont {D.}~\bibnamefont {Popović}},
  \bibinfo {author} {\bibfnamefont {F.}~\bibnamefont {Clerc}}, \bibinfo
  {author} {\bibfnamefont {C.}~\bibnamefont {Koitzsch}}, \bibinfo {author}
  {\bibfnamefont {U.}~\bibnamefont {Probst}}, \bibinfo {author} {\bibfnamefont
  {E.}~\bibnamefont {Bucher}}, \bibinfo {author} {\bibfnamefont
  {H.}~\bibnamefont {Berger}}, \bibinfo {author} {\bibfnamefont
  {D.}~\bibnamefont {Naumović}}, \ and\ \bibinfo {author} {\bibfnamefont
  {P.}~\bibnamefont {Aebi}},\ }\href {\doibase 10.1103/PhysRevB.69.125117}
  {\bibfield  {journal} {\bibinfo  {journal} {Physical Review B}\ }\textbf
  {\bibinfo {volume} {69}} (\bibinfo {year} {2004}),\
  10.1103/PhysRevB.69.125117}\BibitemShut {NoStop}%
\bibitem [{\citenamefont {H.~Shiba}(1986)}]{shiba_phenomenological_1986}%
  \BibitemOpen
  \bibfield  {author} {\bibinfo {author} {\bibfnamefont {K.~N.}\ \bibnamefont
  {H.~Shiba}},\ }in\ \href@noop {} {\emph {\bibinfo {booktitle} {Structural
  {P}hase {T}ransitions in {L}ayered {T}ransition {M}etal {C}ompounds}}},\
  \bibinfo {editor} {edited by\ \bibinfo {editor} {\bibfnamefont
  {K.}~\bibnamefont {Motizuki}}}\ (\bibinfo  {publisher} {D. Reidel Publishing
  Company},\ \bibinfo {year} {1986})\ pp.\ \bibinfo {pages}
  {175--266}\BibitemShut {NoStop}%
\bibitem [{\citenamefont {Mann}\ \emph {et~al.}(2016)\citenamefont {Mann},
  \citenamefont {Baldini}, \citenamefont {Odeh}, \citenamefont {Magrez},
  \citenamefont {Berger},\ and\ \citenamefont {Carbone}}]{mann_probing_2016}%
  \BibitemOpen
  \bibfield  {author} {\bibinfo {author} {\bibfnamefont {A.}~\bibnamefont
  {Mann}}, \bibinfo {author} {\bibfnamefont {E.}~\bibnamefont {Baldini}},
  \bibinfo {author} {\bibfnamefont {A.}~\bibnamefont {Odeh}}, \bibinfo {author}
  {\bibfnamefont {A.}~\bibnamefont {Magrez}}, \bibinfo {author} {\bibfnamefont
  {H.}~\bibnamefont {Berger}}, \ and\ \bibinfo {author} {\bibfnamefont
  {F.}~\bibnamefont {Carbone}},\ }\href {\doibase 10.1103/PhysRevB.94.115122}
  {\bibfield  {journal} {\bibinfo  {journal} {Physical Review B}\ }\textbf
  {\bibinfo {volume} {94}} (\bibinfo {year} {2016}),\
  10.1103/PhysRevB.94.115122}\BibitemShut {NoStop}%
\bibitem [{\citenamefont {Torchinsky}\ and\ \citenamefont
  {Hsieh}(2017)}]{kumar_magnetic_2017}%
  \BibitemOpen
  \bibfield  {author} {\bibinfo {author} {\bibfnamefont {D.~H.}\ \bibnamefont
  {Torchinsky}}\ and\ \bibinfo {author} {\bibfnamefont {D.}~\bibnamefont
  {Hsieh}},\ }in\ \href {\doibase 10.1007/978-3-662-52780-1} {\emph {\bibinfo
  {booktitle} {Magnetic {Characterization} {Techniques} for
  {Nanomaterials}}}},\ \bibinfo {editor} {edited by\ \bibinfo {editor}
  {\bibfnamefont {C.~S.}\ \bibnamefont {Kumar}}}\ (\bibinfo  {publisher}
  {Springer Berlin Heidelberg},\ \bibinfo {address} {Berlin, Heidelberg},\
  \bibinfo {year} {2017})\ pp.\ \bibinfo {pages} {1--40}\BibitemShut {NoStop}%
\bibitem [{\citenamefont {Shen}(1984)}]{shen}%
  \BibitemOpen
  \bibfield  {author} {\bibinfo {author} {\bibfnamefont {Y.}~\bibnamefont
  {Shen}},\ }\href {https://books.google.com/books?id=qYIpAQAAMAAJ} {\emph
  {\bibinfo {title} {The Principles of Nonlinear Optics}}},\ Pure \& Applied
  Optics Series: 1-349\ (\bibinfo  {publisher} {Wiley},\ \bibinfo {year}
  {1984})\BibitemShut {NoStop}%
\bibitem [{\citenamefont {Spijkerman}\ \emph {et~al.}(1997)\citenamefont
  {Spijkerman}, \citenamefont {de~Boer}, \citenamefont {Meetsma}, \citenamefont
  {Wiegers},\ and\ \citenamefont {van Smaalen}}]{spijkerman_x-ray_1997}%
  \BibitemOpen
  \bibfield  {author} {\bibinfo {author} {\bibfnamefont {A.}~\bibnamefont
  {Spijkerman}}, \bibinfo {author} {\bibfnamefont {J.~L.}\ \bibnamefont
  {de~Boer}}, \bibinfo {author} {\bibfnamefont {A.}~\bibnamefont {Meetsma}},
  \bibinfo {author} {\bibfnamefont {G.~A.}\ \bibnamefont {Wiegers}}, \ and\
  \bibinfo {author} {\bibfnamefont {S.}~\bibnamefont {van Smaalen}},\ }\href
  {\doibase 10.1103/PhysRevB.56.13757} {\bibfield  {journal} {\bibinfo
  {journal} {Physical Review B}\ }\textbf {\bibinfo {volume} {56}},\ \bibinfo
  {pages} {13757} (\bibinfo {year} {1997})}\BibitemShut {NoStop}%
\bibitem [{\citenamefont {Berkovic}\ \emph {et~al.}(1989)\citenamefont
  {Berkovic}, \citenamefont {Shen}, \citenamefont {Marowsky},\ and\
  \citenamefont {Steinhoff}}]{berkovic_interference_1988}%
  \BibitemOpen
  \bibfield  {author} {\bibinfo {author} {\bibfnamefont {G.}~\bibnamefont
  {Berkovic}}, \bibinfo {author} {\bibfnamefont {Y.~R.}\ \bibnamefont {Shen}},
  \bibinfo {author} {\bibfnamefont {G.}~\bibnamefont {Marowsky}}, \ and\
  \bibinfo {author} {\bibfnamefont {R.}~\bibnamefont {Steinhoff}},\ }\href
  {\doibase 10.1364/JOSAB.6.000205} {\bibfield  {journal} {\bibinfo  {journal}
  {J. Opt. Soc. Am. B}\ }\textbf {\bibinfo {volume} {6}},\ \bibinfo {pages}
  {205} (\bibinfo {year} {1989})}\BibitemShut {NoStop}%
\bibitem [{\citenamefont {Harter}\ \emph {et~al.}(2015)\citenamefont {Harter},
  \citenamefont {Niu}, \citenamefont {Woss},\ and\ \citenamefont
  {Hsieh}}]{harter_high-speed_2015}%
  \BibitemOpen
  \bibfield  {author} {\bibinfo {author} {\bibfnamefont {J.~W.}\ \bibnamefont
  {Harter}}, \bibinfo {author} {\bibfnamefont {L.}~\bibnamefont {Niu}},
  \bibinfo {author} {\bibfnamefont {A.~J.}\ \bibnamefont {Woss}}, \ and\
  \bibinfo {author} {\bibfnamefont {D.}~\bibnamefont {Hsieh}},\ }\href
  {\doibase 10.1364/OL.40.004671} {\bibfield  {journal} {\bibinfo  {journal}
  {Optics Letters}\ }\textbf {\bibinfo {volume} {40}},\ \bibinfo {pages} {4671}
  (\bibinfo {year} {2015})}\BibitemShut {NoStop}%
\bibitem [{\citenamefont {Torchinsky}\ \emph {et~al.}(2014)\citenamefont
  {Torchinsky}, \citenamefont {Chu}, \citenamefont {Qi}, \citenamefont {Cao},\
  and\ \citenamefont {Hsieh}}]{torchinsky_low_2014}%
  \BibitemOpen
  \bibfield  {author} {\bibinfo {author} {\bibfnamefont {D.~H.}\ \bibnamefont
  {Torchinsky}}, \bibinfo {author} {\bibfnamefont {H.}~\bibnamefont {Chu}},
  \bibinfo {author} {\bibfnamefont {T.}~\bibnamefont {Qi}}, \bibinfo {author}
  {\bibfnamefont {G.}~\bibnamefont {Cao}}, \ and\ \bibinfo {author}
  {\bibfnamefont {D.}~\bibnamefont {Hsieh}},\ }\href {\doibase
  10.1063/1.4891417} {\bibfield  {journal} {\bibinfo  {journal} {Review of
  Scientific Instruments}\ }\textbf {\bibinfo {volume} {85}},\ \bibinfo {pages}
  {083102} (\bibinfo {year} {2014})}\BibitemShut {NoStop}%
\end{thebibliography}%

\end{document}



\title{Supplementary Material for Second Harmonic Generation as a Probe of Broken Mirror Symmetry}


\author{Bryan T. Fichera}
\affiliation{Department of Physics, Massachusetts Institute of Technology, Cambridge, Massachusetts 02139, USA}
\author{Anshul Kogar}
\affiliation{Department of Physics, Massachusetts Institute of Technology, Cambridge, Massachusetts 02139, USA}
\author{Linda Ye}
\affiliation{Department of Physics, Massachusetts Institute of Technology, Cambridge, Massachusetts 02139, USA}
\author{Bilal G{\"o}kce}
\affiliation{Department of Physics, Massachusetts Institute of Technology, Cambridge, Massachusetts 02139, USA}
\affiliation{Technical Chemistry I and Center for Nanointegration Duisburg-Essen (CENIDE), University of Duisburg-Essen, 45141 Essen, Germany}
\author{Alfred Zong}
\affiliation{Department of Physics, Massachusetts Institute of Technology, Cambridge, Massachusetts 02139, USA}
\author{Joseph G. Checkelsky}
\affiliation{Department of Physics, Massachusetts Institute of Technology, Cambridge, Massachusetts 02139, USA}
\author{Nuh Gedik}
\email[]{gedik@mit.edu}
\affiliation{Department of Physics, Massachusetts Institute of Technology, Cambridge, Massachusetts 02139, USA}


\date{\today}


\maketitle

\section{Probing broken mirror symmetry with RA-SHG}
\begin{figure*}
\includegraphics{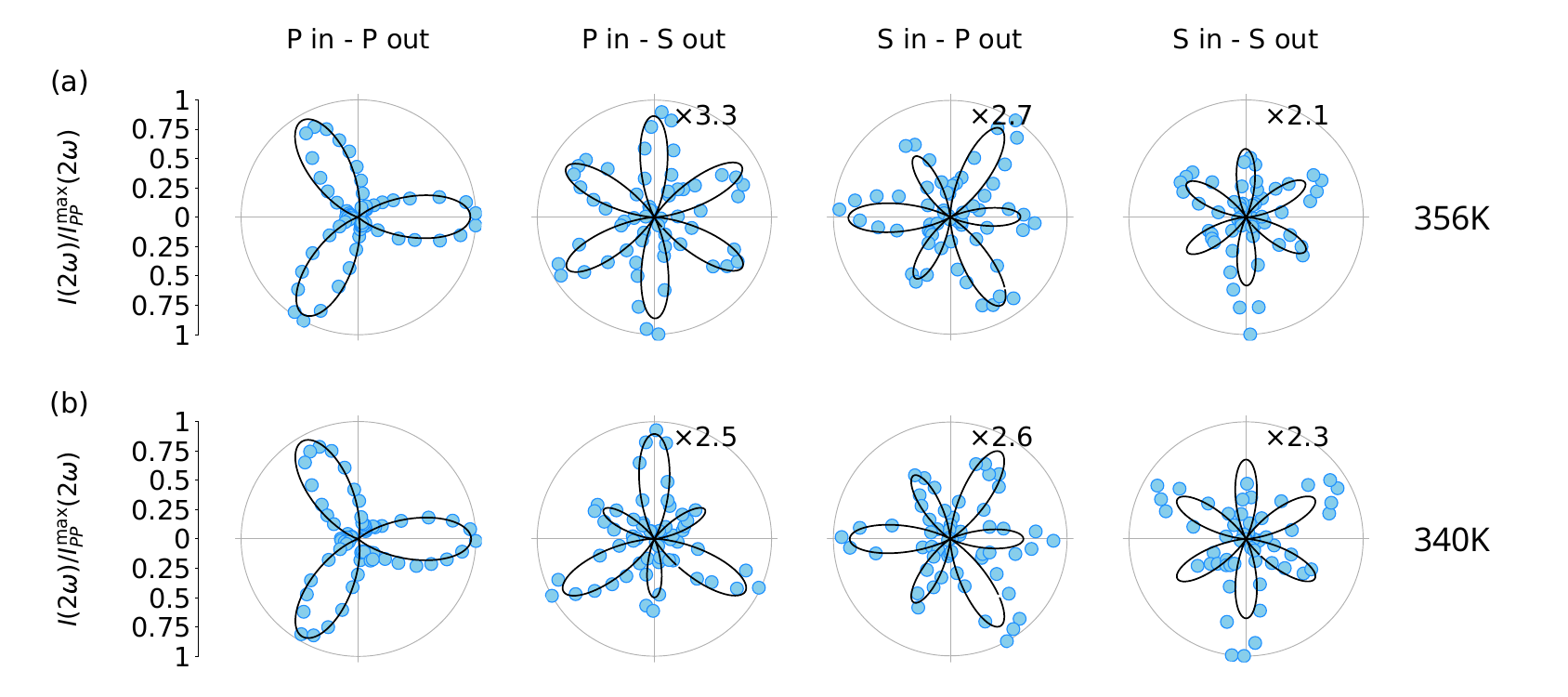}
\caption{\label{fig:figS1}RA-SHG intensity as a function of $\phi$ above (a) and below (b) $T_{IC-NC}=353K$.
Solid lines are are best fits to the data using electric dipole SHG in the surface point groups $C_{3v}$ and $C_3$, respectively.
Data is normalized to the maximum value of the P$_\mathrm{in}$-P$_\mathrm{out}$ signal for each temperature.}
\end{figure*}
\subsection{Complete RA-SHG data\label{sec:complete_a}}
In the first section of the main text, we discuss how RA-SHG is sensitive to the breaking of mirror symmetry in 1$T$-\ce{TaS2}.
In Figs.~\ref{fig:fig1}(a) and \ref{fig:fig1}(b) of the main text, we display RA-SHG data in two polarization channels: P$_\mathrm{in}$-P$_\mathrm{out}$ and P$_\mathrm{in}$-S$_\mathrm{out}$.
For completeness, the other polarization channels are reproduced in Fig.~\ref{fig:figS1}.
Importantly, the fits in the figure only include a surface electric dipole contribution (see main text).
The inability to fit the S$_\mathrm{in}$-S$_\mathrm{out}$ polarization channel with a pure electric dipole contribution suggested to us that a second contribution was necessary in the form of a bulk electric quadrupole term.
This is discussed in the main text as well as in section \ref{sec:Squadrupole}.

\subsection{$\mathscr{I}_{PS}^{(3)}$ as an indicator of broken mirror symmetry\label{sec:Sbmsindicator}}

In the main text, we used the relation given by Eq.~\ref{eq:intensityequation} to show that the breaking of sixfold symmetry in the P$_\mathrm{in}$-S$_\mathrm{out}$ channel is consistent with the lowering of the surface point group from $C_{3v}$ to $C_3$ during the incommensurate (IC) to nearly commensurate (NC) phase transition.
In this section, we show how Eq.~\ref{eq:intensityequation} is derived.

Given any crystallographic point group, it is always possible to compute the form of the susceptibility tensor $\chi^S_{ijk}$.
In $C_3$, the susceptibility tensor is given by~\cite{boyd}
\begin{equation}
\label{eq:Sc3susceptibility}
\chi^S_{ijk} = \begin{pmatrix}
\phantom{-}a & -b & \phantom{-}c\\
-b & -a & -d\\
\phantom{-}c & -d & \phantom{-}0\\
\hline
-b & -a & \phantom{-}d\\
-a & \phantom{-}b & \phantom{-}c\\
\phantom{-}d & \phantom{-}c & \phantom{-}0\\
\hline
\phantom{-}e & \phantom{-}0 & \phantom{-}0\\
\phantom{-}0 & \phantom{-}e & \phantom{-}0\\
\phantom{-}0 & \phantom{-}0 & \phantom{-}f
\end{pmatrix}_{ijk}
\end{equation}
for some $a$, $b$, $c$, $d$, $e$, and $f$ which are dependent on the material.
Here, the threefold axis is taken to be along the $z$ direction.
In the point group $C_{3v}$, the susceptibility tensor is constrained by symmetry so as to take a form which is the same as equation \ref{eq:Sc3susceptibility}, but with $b = d = 0$, i.e.

\begin{equation}
\label{eq:Sc3vsusceptibility}
\chi^S_{ijk} = \begin{pmatrix}
\phantom{-}a & \phantom{-}0 & \phantom{-}c\\
\phantom{-}0 & -a & \phantom{-}0\\
\phantom{-}c & \phantom{-}0 & \phantom{-}0\\
\hline
\phantom{-}0 & -a & \phantom{-}0\\
-a & \phantom{-}0 & \phantom{-}c\\
\phantom{-}0 & \phantom{-}c & \phantom{-}0\\
\hline
\phantom{-}e & \phantom{-}0 & \phantom{-}0\\
\phantom{-}0 & \phantom{-}e & \phantom{-}0\\
\phantom{-}0 & \phantom{-}0 & \phantom{-}f
\end{pmatrix}_{ijk}.
\end{equation}
Here, the mirror planes are taken to be the $xz$-plane and those related to it by rotation about the $z$-axis through $\pm 120^\circ$.

To compute the RA-SHG intensity in the P$_\mathrm{in}$-S$_\mathrm{out}$ channel, we consider a frame of reference in which the plane of incidence is held fixed and the sample is rotated about the optical axis.
This reference frame is equivalent to that of the experimental setup (where the sample is fixed and the plane of incidence rotates), but it is easier to visualize and formulate mathematically.
In this frame, the incoming $P$-polarized light at frequency $\omega$ is parallel to the vector $E_i(\omega) = (-\cos{\theta}, 0, \sin{\theta})^T_i$, and the outgoing $S$-polarized light which we wish to compute is parallel to the $y$-axis.
The RA-SHG intensity in this channel is therefore given by
\begin{equation}
\label{eq:Sipsequation}
I_{PS}^{2\omega}(\phi) \propto \left|P_y^{2\omega}(\phi)\right|^2 \propto \left|\bar{\chi}^S_{yjk}(-\phi)E_j(\omega)E_k(\omega)\right|^2,
\end{equation}
where $\bar{\chi}^S_{ijk}(\phi)$ is the susceptibility tensor corresponding to the sample when it has been rotated by an angle $\phi$.
The relative sign in the argument of $\bar{\chi}^S_{ijk}$ is due to the different reference frame used here compared to Fig.~\ref{fig:fig0}(b), where $\phi$ is defined.

With the rotation matrix
\begin{equation}
R(\phi)_{ij} = \begin{pmatrix}
\cos{\phi} & -\sin{\phi} & 0 \\
\sin{\phi} & \cos{\phi} & 0 \\
0 & 0 & 1
\end{pmatrix}_{ij},
\end{equation}
$\bar{\chi}^S_{ijk}(\phi)$ is given by
\begin{equation}
\label{eq:Srotatechi}
\bar{\chi}^S_{ijk}(\phi) = R(\phi)_{il}R(\phi)_{jm}R(\phi)_{kn}\chi^S_{lmn}.
\end{equation}
Substituting Eq.~\ref{eq:Srotatechi} into Eq.~\ref{eq:Sipsequation}, we then have
\begin{equation}
\label{eq:SintensityequationwithR}
I_{PS}^{2\omega}(\phi) \propto \left|R(-\phi)_{yl}R(-\phi)_{jm}R(-\phi)_{kn}\chi^S_{lmn}E_j(\omega)E_k(\omega)\right|^2.
\end{equation}

With Eq.~\ref{eq:Sc3susceptibility} as $\chi^S_{ijk}$, we arrive at Eq.~\ref{eq:intensityequation} in the main text,
\begin{equation}
\label{eq:Sipsequationfinal}
I_\mathrm{PS}(2\omega) \propto (A_0 + A_1\cos{(3\phi)} + A_2\sin{(3\phi)})^2,
\end{equation}
where
\begin{equation}
\label{eq:sA_0equation}
A_0 = 2d\sin(\theta)\cos(\theta),
\end{equation}
\begin{equation}
\label{eq:sA_1equation}
A_1 = -b\cos^{2}(\theta),
\end{equation}
and
\begin{equation}
\label{eq:sA_2equation}
A_2 = a\cos^{2}(\theta).
\end{equation}
Since $b = d = 0$ in $C_{3v}$, these equations show that $A_0$ and $A_1$ are zero in the presence of mirror symmetry, as cited in the main text.
Furthermore, in $C_{3v}$, Eq.~\ref{eq:Sipsequationfinal} reduces to
\begin{equation}
I_\mathrm{PS}(2\omega) \propto A_2^2\sin^{2}(3\phi),
\end{equation}
which posesses sixfold rotational symmetry.

In the main text, we also remark that the breaking of sixfold rotational symmetry across the IC to NC phase transition can be quantified experimentally by performing a spectral (sine) decomposition of the measured intensity ($I_\mathrm{PS}^{2\omega}(\phi)$) and extracting the third Fourier coefficient, which we called $\mathscr{I}_\mathrm{PS}^{(3)}$.
Here, we define $\mathscr{I}_\mathrm{PS}^{(3)}$ explicitly and show that it is present in 1$T$-\ce{TaS2} only when mirror symmetry is broken.

The spectral decomposition for any given polarization channel ($\Gamma_\mathrm{in}$-$\Gamma_\mathrm{out}$) is defined formally by writing
\begin{equation}
\label{eq:fourierdecomposition}
I_{\Gamma_\mathrm{in}\Gamma_\mathrm{out}}^{2\omega}(\phi) = \sum_{n=0}^{\infty} \mathscr{I}_{\Gamma_\mathrm{in}\Gamma_\mathrm{out}}^{(n)} \sin{\left[n\phi + \psi_{\Gamma_\mathrm{in}\Gamma_\mathrm{out}}^{(n)}\right]}.
\end{equation}
This equation defines $\mathscr{I}_{\Gamma_\mathrm{in}\Gamma_\mathrm{out}}^{(n)}$ and $\psi_{\Gamma_\mathrm{in}\Gamma_\mathrm{out}}^{(n)}$ as the amplitude and phase of the $n$-fold Fourier component of the corresponding SHG intensity. 

Using Eqs.~\ref{eq:Sc3susceptibility},~\ref{eq:SintensityequationwithR}, and~\ref{eq:fourierdecomposition}, we can then compute $\mathscr{I}^{(3)}_\mathrm{PS}$ and $\psi_{PS}^{(3)}$ in the low temperature phase as 
\begin{equation}
\label{eq:i3PS}
\mathscr{I}_{PS}^{(3)}(\chi^S_{ijk}) \propto 4\sqrt{C_1^2+C_2^2}
\end{equation}
and
\begin{equation}
\psi_{PS}^{(3)}(\chi^S_{ijk}) = \atantwo{\left(C_1,C_2\right)},
\end{equation}
where
\begin{equation}
C_1 = -b\cdot d\sin{\theta}\cos^3{\theta},
\end{equation}
\begin{equation}
C_2 = a\cdot d\sin{\theta}\cos^3{\theta},
\end{equation}
$\theta$ is the angle of incidence, and
\begin{equation}
\atantwo{(y, x)} \equiv \begin{cases}
\arctan{\left(\frac{y}{x}\right)} & x >  0 \\
\arctan{\left(\frac{y}{x}\right)}+\pi & y>0,x <  0\\ 
\arctan{\left(\frac{y}{x}\right)}-\pi & y<0,x <  0\\ 
+\frac{\pi}{2} & y>0, x=0\\
-\frac{\pi}{2} & y<0, x=0\\
\mathrm{undefined}& x = y = 0 \end{cases}.
\end{equation}
By comparing equations \ref{eq:Sc3susceptibility} and \ref{eq:Sc3vsusceptibility}, we find that
\begin{equation}
C_1 = C_2 = 0
\end{equation}
in the high temperature phase, so that $\mathscr{I}_{PS}^{(3)}(\chi^S_{ijk}) = 0$. 

Note that Eq.~\ref{eq:i3PS} highlights an important aspect of our experiment, which is that $\mathscr{I}_{PS}^{(3)}$ requires a nonzero angle of incidence to be observed.
This can be understood by noting that $\mathscr{I}_{PS}^{(3)}$ is nonzero only when the tensor element $\chi^S_{yxz}$ is nonzero.
However, for such a term to be observable in the experiment, there needs to be an out-of-plane component to the incoming electric fields, which in turn requires a nonzero angle of incidence.
Moreover, in a system with three-fold symmetry like 1$T$-\ce{TaS2}, the aforementioned tensor element can only be nonzero in a system where mirror symmetry is absent.
To see that this is true, consider the term $P_y(2\omega)=\chi^S_{yxz}E_x(\omega)E_z(\omega)$.
Under an $x \rightarrow -x$ mirror operation, we require that $P_y \rightarrow P_y$, $E_x \rightarrow -E_x$, $E_z \rightarrow E_z$, and $\chi^S_{yxz} \rightarrow \widetilde{\chi}^S_{yxz}$.
However, if this operation is a symmetry of the crystal then we have $\widetilde{\chi}^S_{yxz} = \chi^S_{yxz}$, so that $P_y = -P_y$.
This implies that $\chi^S_{yxz} = 0$.
Therefore both a nonzero angle of incidence and broken mirror symmetry are required for a nonzero $\mathscr{I}_{PS}^{(3)}$ in the geometry of our experiment.

\section{Physical importance of the P$_\mathrm{in}$-S$_\mathrm{out}$ channel}

\begin{figure}
\includegraphics{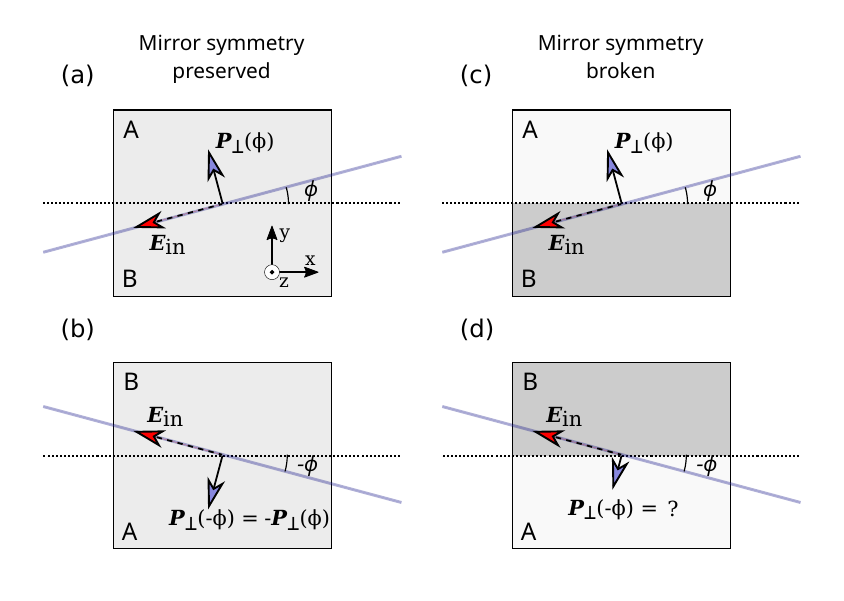}
\caption{\label{fig:mirror}(a) Geometry referenced in showing that the P$_\mathrm{in}$-S$_\mathrm{out}$ geometry is sensitive to the breaking of vertical mirror symmetry. $\bm{E}_\mathrm{in}$ is the input electric field and $\bm{P}_\perp$ is the $S$-polarized component of the polarization. The arrow identifying the direction of $\bm{E}_\mathrm{in}$ is dashed to indicate that there is a component in the $z$-direction (out of the page). The solid blue and dotted black lines represent the plane of incidence and the mirror symmetry plane, respectively, which are in the $z$-direction and make an angle $\phi$ with each other. The two halves (A and B) of the sample are the same color to indicate that mirror symmetry is present in this sample. (b) Mirror image of (a), showing that $\bm{P}_\perp$ flips sign (when measured in the frame of the plane of incidence) under mirror reflection. (c) Same geometry as (a) with broken mirror symmetry, represented by the difference in color between sides A and B. (d) Mirror image of (c).}
\end{figure}

In this section we explain physically why the P$_\mathrm{in}$-S$_\mathrm{out}$ polarization channel is the most sensitive to mirror symmetry breaking in 1$T$-\ce{TaS2}.
Consider the experimental geometry depicted in Fig.~\ref{fig:mirror}(a).
A $P$-polarized electric field $\bm{E}_\mathrm{in}$ is incident on a sample with vertical mirror symmetry about the $xz$ plane (indicated by the sides A and B of the sample being the same color). The geometry is such that the plane of incidence makes an angle $\phi$ with the reflection plane.
Let $\bm{P}_\perp(\phi)$ be the $S$-polarized component of the polarization resulting from second harmonic generation by the interaction of the sample with $\bm{E}_\mathrm{in}$.

Now consider applying the mirror operation both to $\bm{E}_\mathrm{in}$ and to the sample (Fig.~\ref{fig:mirror}(b)).
Then the resulting $S$-polarized component of the polarization (in the frame of the plane of incidence) will be $-\bm{P}_\perp(\phi)$.
However, since the sides A and B in Fig.~\ref{fig:mirror}(b) are the same, by symmetry the problem is equivalent to Fig.~\ref{fig:mirror}(a) but with $\phi$ replaced by $-\phi$.
Thus $\bm{P}_\perp(-\phi) = -\bm{P}_\perp(\phi)$, implying that $\bm{P}_\perp(0) = \bm{P}_\perp(\pi) = 0$.
This along with threefold rotational symmetry is sufficient to prove that the RA-SHG pattern from IC-phase 1$T$-\ce{TaS2} in the P$_\mathrm{in}$-S$_\mathrm{out}$ polarization channel is purely sixfold-symmetric.
However, in the case where mirror symmetry is broken (Figs.~\ref{fig:mirror}(c-d)) the above argument no longer holds and the constraint $\bm{P}_\perp(-\phi) = -\bm{P}_\perp(\phi)$ is relaxed.
Therefore in 1$T$-\ce{TaS2}, $\mathscr{I}_\mathrm{PS}^{(3)}$ is allowed in the NC phase but not in the IC phase.

\renewcommand{\arraystretch}{1.5}
\begin{table*}[t]
\begin{tabular}{c|c|p{8cm}}
\hline
\hline
\textbf{Initial group} & \textbf{Final group} & \multicolumn{1}{>{\centering\arraybackslash}m{100mm}}{\textbf{Indicator(s)}}\\
\hline
$T_{d}$ & $T$ & None \\
$T_{d}$ & $D_{2d}$ & None \\
$T_{d}$ & $C_{3v}$ & None \\
$C_{6v}$ & $C_{3v}$ & $\mathscr{I}_{PP}^{(3)}$, $\mathscr{I}_{PP}^{(6)}$, $\mathscr{I}_{PS}^{(0)}$, $\mathscr{I}_{PS}^{(6)}$, $\mathscr{I}_{SP}^{(3)}$, $\mathscr{I}_{SP}^{(6)}$, $\mathscr{I}_{SS}^{(0)}$, $\mathscr{I}_{SS}^{(6)}$ \\
$C_{6v}$ & $C_{2v}$ & $\mathscr{I}_{PP}^{(2)}$, $\mathscr{I}_{PP}^{(4)}$, $\mathscr{I}_{PS}^{(0)}$, $\mathscr{I}_{PS}^{(4)}$, $\mathscr{I}_{SP}^{(2)}$, $\mathscr{I}_{SP}^{(4)}$ \\
$C_{6v}$ & $C_{6}$ & $\mathscr{I}_{PS}^{(0)}$ \\
$C_{4v}$ & $C_{2v}$ & $\mathscr{I}_{PP}^{(2)}$, $\mathscr{I}_{PP}^{(4)}$, $\mathscr{I}_{PS}^{(0)}$, $\mathscr{I}_{PS}^{(4)}$, $\mathscr{I}_{SP}^{(2)}$, $\mathscr{I}_{SP}^{(4)}$ \\
$C_{4v}$ & $C_{4}$ & $\mathscr{I}_{PS}^{(0)}$ \\
$D_{3h}$ & $C_{3h}$ & $\psi_{PP}^{(6)}$$^*$, $\psi_{PS}^{(6)}$$^*$, $\psi_{SP}^{(6)}$$^*$, $\psi_{SS}^{(6)}$$^*$ \\
$D_{3h}$ & $C_{2v}$ & None \\
$D_{3h}$ & $D_{3}$ & $\mathscr{I}_{PS}^{(3)}$ \\
$C_{3h}$ & $C_{3}$ & $\psi_{PP}^{(1)}$$^*$, $\psi_{PP}^{(2)}$$^*$, $\psi_{PP}^{(3)}$$^*$, $\psi_{PP}^{(4)}$$^*$, $\psi_{PP}^{(5)}$$^*$, $\psi_{PP}^{(6)}$$^*$, $\psi_{PS}^{(1)}$$^*$, $\psi_{PS}^{(2)}$$^*$, $\psi_{PS}^{(3)}$$^*$, $\psi_{PS}^{(4)}$$^*$, $\psi_{PS}^{(5)}$$^*$, $\psi_{PS}^{(6)}$$^*$, $\psi_{SP}^{(1)}$$^*$, $\psi_{SP}^{(2)}$$^*$, $\psi_{SP}^{(3)}$$^*$, $\psi_{SP}^{(4)}$$^*$, $\psi_{SP}^{(5)}$$^*$, $\psi_{SP}^{(6)}$$^*$, $\psi_{SS}^{(2)}$$^*$, $\psi_{SS}^{(4)}$$^*$, $\psi_{SS}^{(6)}$$^*$ \\
$C_{3v}$ & $C_{3}$ & $\psi_{PP}^{(3)}$, $\psi_{PP}^{(6)}$$^*$, $\mathscr{I}_{PS}^{(3)}$, $\psi_{PS}^{(6)}$$^*$, $\psi_{SP}^{(3)}$, $\psi_{SP}^{(6)}$$^*$, $\psi_{SS}^{(6)}$$^*$ \\
$C_{3v}$ & $C_{1h}$ & $\mathscr{I}_{PP}^{(1)}$, $\mathscr{I}_{PP}^{(2)}$, $\mathscr{I}_{PP}^{(4)}$, $\mathscr{I}_{PP}^{(5)}$, $\mathscr{I}_{PS}^{(1)}$, $\mathscr{I}_{PS}^{(2)}$, $\mathscr{I}_{PS}^{(3)}$, $\mathscr{I}_{PS}^{(4)}$, $\mathscr{I}_{PS}^{(5)}$, $\mathscr{I}_{SP}^{(1)}$, $\mathscr{I}_{SP}^{(2)}$, $\mathscr{I}_{SP}^{(4)}$, $\mathscr{I}_{SP}^{(5)}$, $\mathscr{I}_{SS}^{(2)}$, $\mathscr{I}_{SS}^{(4)}$ \\
$D_{2d}$ & $S_{4}$ & $\psi_{PP}^{(4)}$$^*$, $\psi_{PS}^{(4)}$$^*$, $\psi_{SP}^{(4)}$$^*$ \\
$D_{2d}$ & $D_{2}$ & $\mathscr{I}_{PS}^{(2)}$ \\
$C_{2v}$ & $C_{2}$ & $\psi_{PP}^{(2)}$$^*$, $\psi_{PP}^{(4)}$$^*$, $\mathscr{I}_{PS}^{(2)}$, $\psi_{PS}^{(4)}$$^*$, $\psi_{SP}^{(2)}$$^*$, $\psi_{SP}^{(4)}$$^*$ \\
$C_{2v}$ & $C_{1h}$ & $\mathscr{I}_{PP}^{(1)}$, $\mathscr{I}_{PP}^{(3)}$, $\mathscr{I}_{PP}^{(5)}$, $\mathscr{I}_{PP}^{(6)}$, $\mathscr{I}_{PS}^{(1)}$, $\mathscr{I}_{PS}^{(2)}$, $\mathscr{I}_{PS}^{(3)}$, $\mathscr{I}_{PS}^{(5)}$, $\mathscr{I}_{PS}^{(6)}$, $\mathscr{I}_{SP}^{(1)}$, $\mathscr{I}_{SP}^{(3)}$, $\mathscr{I}_{SP}^{(5)}$, $\mathscr{I}_{SP}^{(6)}$, $\mathscr{I}_{SS}^{(0)}$, $\mathscr{I}_{SS}^{(2)}$, $\mathscr{I}_{SS}^{(4)}$, $\mathscr{I}_{SS}^{(6)}$ \\
\hline\multicolumn{3}{l}{$^*$\footnotesize{ Measured from $\pi/2$}}\\
\end{tabular}
\caption{Indicators of broken mirror symmetry in mirror-symmetry-breaking transitions between noncentrosymmetric point groups. Values in the rightmost column (defined in Eq.~\ref{eq:fourierdecomposition}) are zero in the initial group and nonzero in the final group. In all cases, the analysis was done in a geometry such that the sample normal is parallel to the broken mirror plane (i.e. the mirror plane is vertical).}
\label{tab:indicators}
\end{table*}

In contrast to the $S$-polarized output ($\bm{P}_\perp(\phi)$), the $P$-polarized output ($\bm{P}_\parallel(\phi)$) does not change sign (as defined in the frame of the plane of incidence) under mirror reflection.
Therefore the constraint for $P$-polarized output is that $\bm{P}_\parallel(\phi) = \bm{P}_\parallel(-\phi)$.
Importantly, this constraint does not forbid fourier components like $\mathscr{I}_\mathrm{PP}^{(3)}$ and $\mathscr{I}_\mathrm{SP}^{(3)}$, even when mirror symmetry is present.

Finally, while the $S_\mathrm{in}$-$S_\mathrm{out}$ channel is subject to the constraint $\bm{P}_\perp(-\phi) = -\bm{P}_\perp(\phi)$, it can also be shown (see section~\ref{sec:sstwofold}) that in the electric dipole regime, $S_\mathrm{in}$-$S_\mathrm{out}$ is twofold-symmetric regardless of the crystallographic point group.
Therefore the $P_\mathrm{in}$-$S_\mathrm{out}$ polarization channel is especially sensitive to the breaking of vertical mirror plane symmetry in the geometry depicted in Fig.~\ref{fig:mirror}.

\section{Generalization to other mirror symmetry-breaking transitions}

While the argument above is effective in the case of a $C_{3v}$ to $C_3$ transition (as in 1$T$-\ce{TaS2}), the presence or absence of other symmetries in different point groups may call for symmetry arguments which differ from those above.
Therefore, to generalize this work we provide in Table~\ref{tab:indicators} a list of mirror-symmetry-breaking transitions between noncentrosymmetric point groups and identify corresponding terms (like $\mathscr{I}_\mathrm{PS}^{(3)}$ in 1$T$-\ce{TaS2}) which adopt nonzero values in the low-symmetry phase.
This table was generated using the same analysis as in section~\ref{sec:Sbmsindicator} for 1$T$-\ce{TaS2}. 
Note that while RA-SHG is only nonzero in noncentrosymmetric point groups, it can still be used in centrosymmetric materials because crystal surfaces necessarily break inversion symmetry.

In most of the transitions in Table~\ref{tab:indicators}, $\mathscr{I}_{PS}^{(n)}$ can be used as an indicator of broken mirror symmetry for some $n$.
However, for others the RA-SHG pattern rotates rather than (or in addition to) changing symmetry, which is indicated by a change in the phase $\psi^{(n)}_{\Gamma_\mathrm{in}\Gamma_\mathrm{out}}$ of certain Fourier components.
Additionally, it should be observed that for a minority of the transitions in Table~\ref{tab:indicators}, RA-SHG is not sensitive to the breaking of mirror symmetry (e.g. in the transition from $T_d$ to $T$, two point groups for which the susceptibility tensor is equivalent).

\section{Probing the sense of planar chirality with RA-SHG}
\subsection{Complete RA-SHG data\label{sec:complete_b}}
\begin{figure*}
\includegraphics{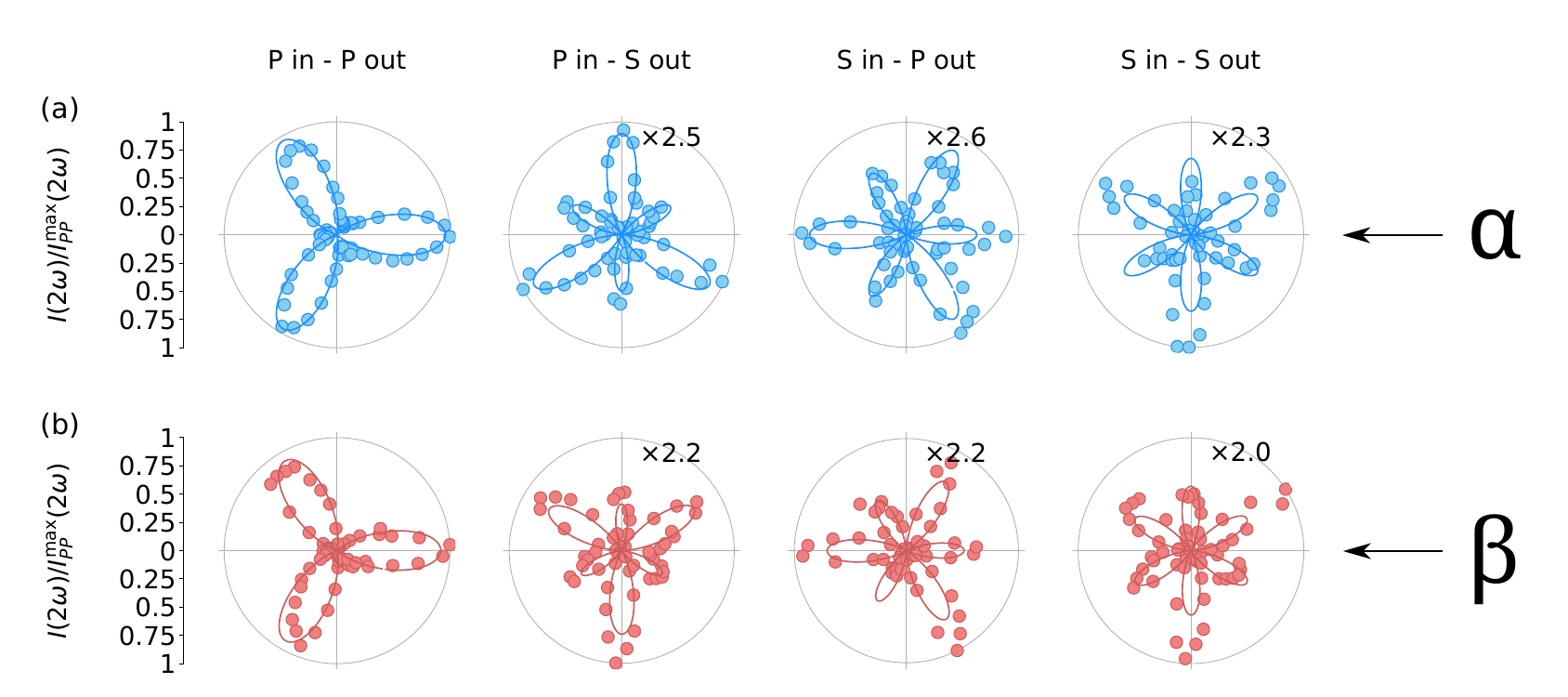}
\caption{\label{fig:figS2}Second harmonic intensity as a function of $\phi$ from two samples of 1$T$-\ce{TaS2} in the NC phase ($T=340K$).
The labels $\alpha$ and $\beta$ refer to the two degenerate mirror-image configurations which can exist in the NC phase.
The solid line in (a) is a fit to the data using the surface point group $C_3$.
The fit in (b) was generated by performing a mirror operation (see section \ref{sec:mirrorflip}) to the numerical susceptibility tensor obtained from (a).
Data is normalized to the maximum value in the P$_\mathrm{in}$-P$_\mathrm{out}$ polarization channel for each sample.}
\end{figure*}

As in section \ref{sec:complete_a}, for completeness we reproduce in Fig.~\ref{fig:figS2} the RA-SHG data in the $\alpha$ and $\beta$ charge density wave (CDW) configurations, including the two polarization channels which were truncated from Fig.~\ref{fig:fig2} in the main text.
The poor fit in the S$_\mathrm{in}$-S$_\mathrm{out}$ polarization channel again indicated to us that there was an additional bulk quadrupole contribution to the signal.

\subsection{Electron diffraction confirming the sample is single domain}
\begin{figure}
\includegraphics{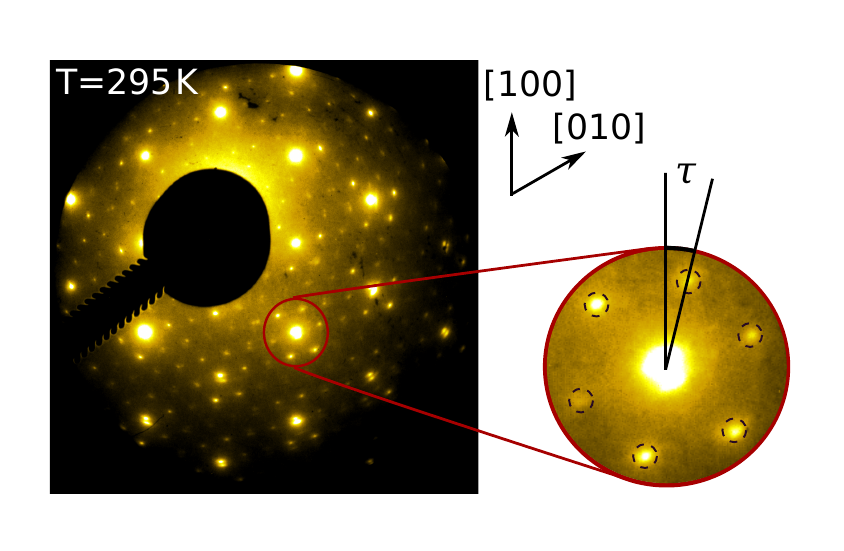}
\caption{\label{fig:diffraction}Electron diffraction data from a sample of 1$T$-\ce{TaS2} in the NC phase of the same batch as the one used in RA-SHG measurements.
Surrounding each Bragg reflection can be seen six CDW peaks, which are rotated about $\tau=13^\circ$ from the high-symmetry axes.
The sign of the rotation angle $\tau$ indicates whether the CDW configuration is $\alpha$ or $\beta$.
The lack of additional CDW peaks $-13^\circ$ from the high-symmetry axes indicates that the sample is single-domain.
}
\end{figure}

To confirm that 1$T$-\ce{TaS2} is single-domain in the NC phase, we performed electron diffraction on a sample from the same batch as the one used for the RA-SHG measurements.
The absence of CDW sister peaks in Fig.~\ref{fig:diffraction} demonstrates that the sample is single-domain, in agreement with previous reports~\cite{zong_ultrafast_2018, wilson_charge-density_1975, bovet_pseudogapped_2004, shiba_phenomenological_1986}.
The thickness of the sample measured by electron diffraction was $\sim$80$\si{nm}$, which is approximately equal to the penetration depth of 1$T$-\ce{TaS2} at $800\si{nm}$~\cite{mann_probing_2016}.
We note that recent studies have demonstrated that it is possible to inject mirror domains into an otherwise uniform sample of 1$T$-\ce{TaS2} using intense pulses of light~\cite{zong_ultrafast_2018}.
To control for this effect, we used an incident fluence of 1.4 \si{mJ/cm^2} which is well below the threshold fluence 5 \si{mJ/cm^2} reported in Ref.~\onlinecite{zong_ultrafast_2018}.

\subsection{Fitting RA-SHG data in the $\alpha$ and $\beta$ configurations}
\label{sec:mirrorflip}
To fit the data in Fig.~\ref{fig:fig2} of the main text, we performed the following procedure.
First, we used the six independent elements in the $C_{3}$ susceptibility tensor given by equation \ref{eq:Sc3susceptibility} as fitting parameters to fit the data on the $\alpha$ sample (Fig.~\ref{fig:figS2}(a)) to 
\begin{equation}
|P_i(\phi)|^2 \propto |R(-\phi)_{il}R(-\phi)_{jm}R(-\phi)_{kn}\chi_{lmn}^SE_j(\omega)E_k(\omega)|^2.
\end{equation}
The fit was constrained in such a way as to forbid large changes in the susceptibility elements that exist in both phases (i.e. $a$, $c$, $e$, and $f$).
This gave us a set of six numbers $\{\chi_{ijk}^\alpha\}$, from which we can form the tensor $\chi_{ijk}^\alpha$.
The cleaving operation can be represented by the operator $\mathcal{C} = A\Gamma(\gamma)R_x$, where $R_x$ is the operator which rotates the sample $180^\circ$ about the $x$-axis,
\begin{equation}
\label{eq:chiralityswitcher}
R_x = \begin{pmatrix}
1 & 0 & 0 \\
0 & -1 & 0 \\
0 & 0 & -1 \\
\end{pmatrix},
\end{equation}
$\Gamma(\gamma)$ is the operator which rotates the sample about the $z$-axis by an arbitrary angle $\gamma$,
\begin{equation}
\Gamma(\gamma) = \begin{pmatrix}
\cos{\gamma} & -\sin{\gamma} & 0 \\
\sin{\gamma} & \cos{\gamma} & 0 \\
0 & 0 & 1
\end{pmatrix},
\end{equation}
and $A$ is a positive overall constant which represents minor, day-to-day fluctuations in experimental conditions.
If we were able to cleave the sample exactly along the high-symmetry axis, then $\gamma$ would be $0^\circ$ and $\Gamma(\gamma)$ would be identity.
Also note that the fact that the determinant of $\mathcal{C}$ in the $(x, y)$ subspace is $-1$ is equivalent to the statement that $\mathcal{C}$ switches the planar chirality of the sample.

To simulate the effect of cleaving the sample, we therefore applied $\mathcal{C}$ to $\chi_{ijk}^\alpha$ to form $\chi_{ijk}^\beta(\{\chi_{ijk}^\alpha\}, \gamma, A)$.
Formally, this amounts to computing
\begin{equation}
\label{eq:SapplyingC}
\chi_{ijk}^\beta(\{\chi_{ijk}^\alpha\}, \gamma, A) = \mathcal{C}_{il}(\gamma, A)\mathcal{C}_{jm}(\gamma, A)\mathcal{C}_{kn}(\gamma, A)\chi_{lmn}^\alpha.
\end{equation}

Now, $\chi_{ijk}^\beta(\{\chi_{ijk}^\alpha\}, \gamma, A)$ is a function of only two free parameters ($\gamma$ and $A$).
We find that with the proper choice of $\gamma$ and $A$, the signal computed with $\chi_{ijk}^\beta$ collapses onto the data in Fig.~\ref{fig:figS2}(b).

By applying $R_x$ to the tensor given in Eq.~\ref{eq:Sc3susceptibility}, one can see that the effect of cleaving the sample is to flip the sign of $b$ and $d$.
By examing Eqs.~\ref{eq:Sipsequationfinal}-\ref{eq:sA_2equation}, and noting that $A_1\approx 0$ within the resolution of our instrument, we see that this reproduces the change of the orientation of the RA-SHG pattern in the P$_\mathrm{in}$-S$_\mathrm{out}$ channel depicted in Fig.~\ref{fig:fig2} of the main text.

%

\section{Bulk broken mirror symmetry in 1$T$-\ce{TaS2}} \label{sec:Squadrupole}

\subsection{Complete RA-SHG data at $T=295K$}

\begin{figure*}
\includegraphics{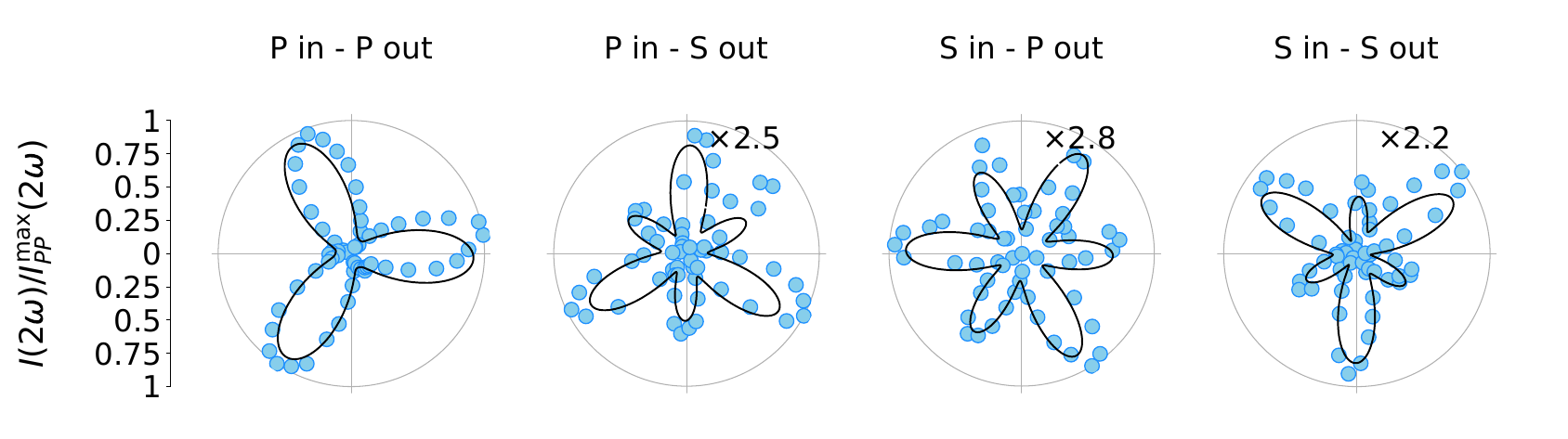}
\caption{\label{fig:295K}RA-SHG intensity as a function of $\phi$ at $T=295K$.
Solid lines are best fits to the data using a surface electric dipole term in the point group $C_3$, as well as a bulk electric quadrupole term in the point group $S_6$.
Data is normalized to the maximum value of the P$_\mathrm{in}$-P$_\mathrm{out}$ signal.}
\end{figure*}

\begin{figure*}
\includegraphics{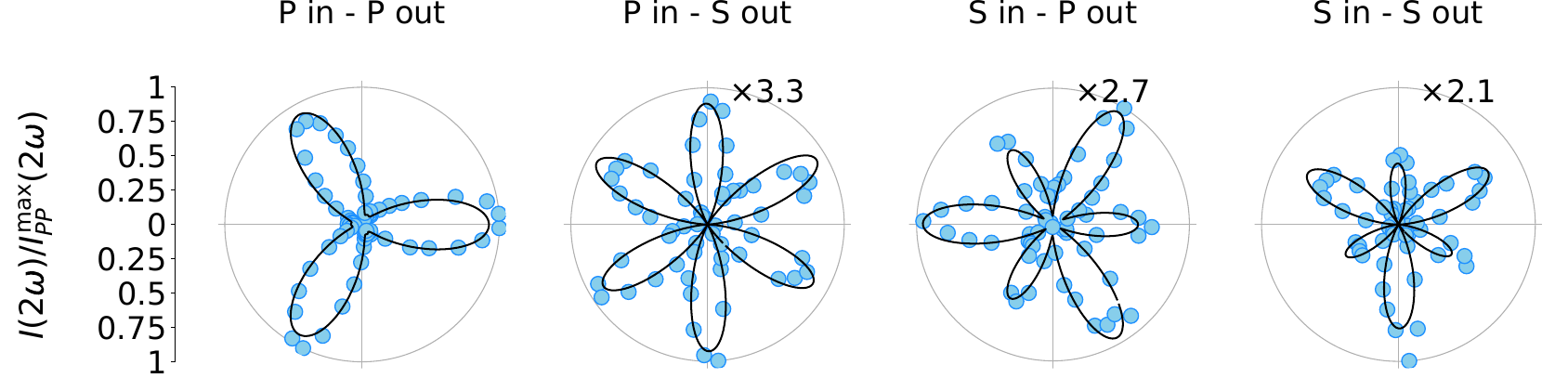}
\caption{\label{fig:356K}RA-SHG intensity as a function of $\phi$ at $T=356K$.
Solid lines are best fits to the data using a surface electric dipole term in the point group $C_{3v}$, as well as a bulk electric quadrupole term in the point group $S_6$.
Data is normalized to the maximum value of the P$_\mathrm{in}$-P$_\mathrm{out}$ signal.}
\end{figure*}

As in section~\ref{sec:complete_b}, for completeness we reproduce in Fig.~\ref{fig:295K} the RA-SHG data at $T=295K$ (NC phase) which is used in Fig.~\ref{fig:fig3} to argue that there is a nonzero bulk electric quadrupole component to the measured signal.
Furthermore, we reproduce in Fig.~\ref{fig:356K} the RA-SHG data at $T=356K$ (IC phase) to show that we can fit all four polarization combinations using the surface point group $C_{3v}$ as well as a bulk point group $S_6$, which is the highest-symmetry subgroup of the undistorted point group which breaks mirror symmetry (see below).
The S$_\mathrm{in}$-S$_\mathrm{out}$ channel of Fig.~\ref{fig:356K} is shown in Fig.~\ref{fig:fig3} of the main text.

\subsection{Symmetry analysis of RA-SHG data} \label{sec:sstwofold}

In the last section of the main text, we used the relation given by Eq.~\ref{eq:qintensityequation} to argue that the breaking of sixfold symmetry in the S$_\mathrm{in}$-S$_\mathrm{out}$ polarization channel indicates that bulk mirror symmetry is broken in the IC phase of 1$T$-\ce{TaS2}.
Here we derive Eq.~\ref{eq:qintensityequation} as we did Eq.~\ref{eq:intensityequation} in section~\ref{sec:Sbmsindicator}.

As mentioned in the main text, the effective polarization for bulk electric quadrupole SHG is given by~\cite{kumar_magnetic_2017, shen}
\begin{equation}
\nabla_j Q_{ij} = 2i\chi_{ijkl}^Qk_j E_k E_l,
\end{equation}
so that the total intensity in the S$_\mathrm{in}$-S$_\mathrm{out}$ channel (including surface electric dipole) is given by
\begin{equation}
\label{eq:SQintensity}
I_{SS}^{2\omega}(\phi) \propto \left|\left(\bar{\chi}^S_{yjk}(-\phi)+2i \bar{\chi}_{ypjk}^Q(-\phi)k_p\right)E_j(\omega)E_k(\omega)\right|^2,
\end{equation}
where
\begin{equation}
\bar{\chi}^S_{ijk}(\phi) = R(\phi)_{il}R(\phi)_{jm}R(\phi)_{kn}\chi^S_{lmn},
\end{equation}
\begin{equation}
\bar{\chi}^Q_{ipjk}(\phi) = R(\phi)_{il}R(\phi)_{pq}R(\phi)_{jm}R(\phi)_{kn}\chi_{lqmn}^Q,
\end{equation}
and $E_i(\omega) = (0, 1, 0)^T_i$.

In the NC phase of 1$T$-\ce{TaS2}, the appropriate assignment for the surface and bulk point groups is given by diffraction measurements~\cite{spijkerman_x-ray_1997} to be $C_3$ and $S_6$, respectively.
Therefore for $\chi^S_{ijk}$ we use the tensor defined by Eq.~\ref{eq:Sc3susceptibility}, and for $\chi^Q_{ijkl}$ we use the tensor given in Ref.~\onlinecite{boyd} for the point group $S_6$ (which for the sake of brevity we do not reproduce here).
With these tensors, we then have
\begin{equation}
\label{eq:Sequationwithbs}
I_\mathrm{SS}(2\omega) \propto (B_0+B_1\cos{(3\phi)}+B_2\sin{(3\phi)})^2,
\end{equation}
with
\begin{equation}
B_0 = \chi^Q_{xyxx}\sin{\theta},
\end{equation}
\begin{equation}
B_1 = \chi^S_{yyy} -\chi^Q_{xzxy}\cos{\theta},
\end{equation}
and
\begin{equation}
B_2 = -\chi^S_{xxx}+\chi^Q_{xzyy}\cos{\theta}.
\end{equation}

Importantly, if bulk mirror symmetry were restored in the IC phase we would have $\chi^Q_{xyxx} = \chi^Q_{xzxy} = 0$, and therefore $B_0=0$. 
This can be checked by inspecting the quadrupole susceptibility tensor for the undistorted bulk point group $D_{3d}$ (which preserves vertical mirror symmetry) defined in Ref.~\onlinecite{boyd}.

Since $B_0=0$ if the pattern possesses sixfold rotational symmetry (as can be seen by taking $\phi\to\phi+60^\circ$), we have that the breaking of sixfold rotational symmetry in the RA-SHG pattern (Fig.~\ref{fig:fig3}(c) of the main text) implies $B\neq 0$.
This implies that $\chi^Q_{xyxx} \neq 0$, and therefore mirror symmetry is broken in the bulk.

There are two additional important remarks which can be made about Eq.~\ref{eq:Sequationwithbs}.
For one, it should be noted that the $\theta$ dependence observed in Figs.~\ref{fig:fig3}(a) and~\ref{fig:fig3}(b) is consistent with Eq.~\ref{eq:Sequationwithbs}, since $B_0=0$ for $\theta=0$.
Additionally, we see that in the absence of the quadrupole component (i.e. for purely electric dipole SHG), $B_0 = 0$ and the pattern should have sixfold rotational symmetry.
Since the data in Figs.~\ref{fig:fig3}(a) and~\ref{fig:fig3}(c) clearly break this symmetry, this supports our claim that their is an extra contribution to the RA-SHG beyond the surface electric dipole contribution.

A more intuitive way to understand why the threefold spectral component in Figs.~\ref{fig:fig3}(a) and~\ref{fig:fig3}(c) is not allowed by pure electric dipole SHG is to note that, for purely electric dipole SHG, the rotational anisotropy always has at least twofold symmetry in the S$_\mathrm{in}$-S$_\mathrm{out}$ channel.
To see that this is true, consider the effect of taking $\phi\rightarrow\phi+\pi$ in an $S$-polarized input geometry.
In the frame where the sample is stationary, this simply changes the sign of the input field. 
Since the polarization is proportional to the square of the input field ($P_i(2\omega) = \chi_{ijk}E_j E_k$), then it will be symmetric under $\phi\rightarrow\phi+\pi$, as will be its projection out of the plane of incidence (note, however, that because the component of $\bm{P}(2\omega)$ parallel to the direction of propogation is not visible in the experiment, the same cannot be said for the $P$-polarized output).
Therefore, if the measured RA-SHG in the S$_\mathrm{in}$-S$_\mathrm{out}$ channel lacks twofold symmetry, then there is a contribution to the rotational anisotropy which exists beyond the electric dipole.

For the sake of completeness, we also note here two alternative explanations for the breaking of sixfold rotational symmetry depicted in Fig.~\ref{fig:fig3}(c) which do not appeal to an additional quadrupole contribution.
Firstly, it is possible that the surface of the sample was rough or nonuniform in such a way as to modulate the observed SHG intensity as a function of $\phi$.
In principle, this could result in a nonzero $B_0$.
To account for this artifact, we measured SHG on 1$T$-\ce{TaS2} in multiple locations and on multiple different samples, and found that the results in this work were consistent across all measurements.
Furthermore, with SHG measurements there is always the possibility of an additional contribution arising from the presence of adsorbates on the sample surface\cite{berkovic_interference_1988}.
While such a contribution may be present in our experiment, we note that this too is an unlikely explanation for the nonzero $B_0$ because of the wavevector dependence of Figs.~\ref{fig:fig3}(a) and (b).
This is because the dominant contribution from adsorbates is generally of the electric dipole type and therefore should not depend on $\bm{k}$ in S$_\mathrm{in}$-S$_\mathrm{out}$.

\section{Methods}
RA-SHG data was taken with the 800\si{\nm} pulsed output of a regeneratively amplified Ti:Sapphire laser operating at 5kHz.
The beam was scattered by a transmissive phase grating into multiple diffraction orders, and the +1-order component was steered through a polarizer and focused onto the sample with a fluence of 1.4 \si{mJ/cm^2} and a spot diameter of $\sim$150$\mu$m at a $10^\circ$ angle with respect to the $(001)$ sample normal.
This experimental geometry is similar to that described in Refs.~\onlinecite{harter_high-speed_2015} and \onlinecite{torchinsky_low_2014}.
The reflected light at 400\si{\nm} was directed through an analyzer and into a photomultiplier tube, where the intensity was measured with a lock-in amplifier synchronized to the 5kHz pulsed output of the laser with a 1\si{\ms} time constant.
We rotated the phase grating at $\sim$5Hz and recorded the signal as a function of time with an oscilloscope triggered on an optical rotary encoder marking $360^\circ$ rotations.
Each polarization combination required averaging 5000 rotations.
Using the aforementioned encoder, we expressed the corresponding time trace in terms of the azimuthal angle $\phi$ (Fig.~\ref{fig:fig0}(b)) to obtain the full rotational anisotropy.

The thickness of the samples used in the RA-SHG experiments was $\sim$50$\si{\mu m}$.

\bibliography{NCtas2sup}

\makeatletter\@input{xx.tex}\makeatother